\documentclass[5p,times]{elsarticle}

\usepackage{amsmath}
\usepackage{soul}
\usepackage{color}
\usepackage[usenames,dvipsnames]{xcolor}
\usepackage{multirow}
\usepackage{longtable}
\usepackage{graphicx}
\usepackage{longtable}
\usepackage{tikz}
\usepackage{amsmath}

\usepackage{booktabs} 
\usepackage{longtable}
\usepackage{xurl}

\PassOptionsToPackage{hyphens}{url}\usepackage{hyperref}

\journal{Journal of \LaTeX\ Templates}

\begin{document}

\begin{frontmatter}
\title{MalDetConv: Automated Behaviour-based Malware Detection Framework Based on Natural Language Processing and Deep Learning Techniques}
\author{Pascal Maniriho}
\author{Abdun Naser Mahmood}
\author{Mohammad Jabed Morshed Chowdhury}
\address{Department of Computer Science and Information Technology, La Trobe University, Melbourne, VIC, Australia}




\begin{abstract}
The popularity of Windows attracts the attention of hackers/cyber-attackers, making Windows devices the primary target of malware attacks in recent years. Several sophisticated malware variants and anti-detection methods have been significantly enhanced and as a result, traditional malware detection techniques have become less effective. This work presents MalBehavD-V1, a new behavioural dataset of Windows Application Programming Interface (API) calls extracted from benign and malware executable files using the dynamic analysis approach. In addition, we present MalDetConV, a new automated behaviour-based framework for detecting both existing and zero-day malware attacks. MalDetConv uses a text processing-based encoder to transform features of API calls into a suitable format supported by deep learning models. It then uses a hybrid of convolutional neural network (CNN) and bidirectional gated recurrent unit (CNN-BiGRU) automatic feature extractor to select high-level features of the API Calls which are then fed to a fully connected neural network module for malware classification. MalDetConv also uses an explainable component that reveals features that contributed to the final classification outcome, helping the decision-making process for security analysts. The performance of the proposed framework is evaluated using our MalBehavD-V1 dataset and other benchmark datasets. The detection results demonstrate the effectiveness of MalDetConv over the state-of-the-art techniques with detection accuracy of 96.10\%, 95.73\%, 98.18\%, and 99.93\% achieved while detecting unseen malware from MalBehavD-V1, Allan and John, Brazilian, and Ki-D datasets, respectively. The experimental results show that MalDetConv is highly accurate in detecting both known and zero-day malware attacks on Windows devices.

\end{abstract}

\begin{keyword}
\texttt{} Malware \sep dynamic analysis \sep malware detection \sep Convolutional neural network\sep API Calls \sep word embedding \sep machine learning \sep deep learning 
\end{keyword}

\end{frontmatter}

\section{Introduction}
\label{introduc}
As Internet-based applications continue to shape various businesses around the globe, malware threats have become a severe problem for computing devices such as desktop computers, smartphones, local servers, remote servers, and IoT devices. It is expected that by 2023 \cite{cisco2020cisco} the total number of devices connected to IP networks will be around 29.3 billion, while predictions show that internet of things (IoT) devices in use will be more than 29 billion in 2030 \cite{IoTconn31:online}, resulting in a massive interconnection of various networked devices. As the number of connected devices continues to rise exponentially, this has also become a motivating factor for cyber-criminals to develop new advanced malware programs that disrupt, steal sensitive data, damage, and exploit various vulnerabilities. The widespread use of different malware variants makes the existing security systems ineffective whereby, millions of devices are infected by various forms of malware such as worms, ransomware, backdoors, computer viruses, and Trojans\cite{maniriho2021study} \cite{ANotSoCo81:online}. Accordingly, there has been a significant increase of new malware targeting Windows devices, i.e., the number of malware samples increased by 23\% (9.5 million) \cite{Over100M58:online} from 2020 to 2021. About 107.27 million new malware samples were created to compromise windows devices in 2021, showing an increase of 16.53 million samples over 2020 with an average of 328,073 malware samples produced daily \cite{Over100M58:online}.

The application of signature-based malware detection systems such as anti-virus programs that rely on a database of signatures extracted from the previously identified malware samples is prevalent. In static malware  analysis, signatures are malware's unique identities which are extracted from malware without executing the suspicious program \cite{zhang2019static} \cite{naik2021fuzzy}.

Some of the static-based malware analysis techniques were implemented using printable strings, opcode sequences, and static API calls\cite{Singh2020a},  \cite{sun2019opcode} \cite{huda2016hybrids}. As signature-based systems rely on previously seen signatures to detect malware threats, they have become ineffective due to a huge number of new malware variants coming out every day \cite{5615097}. Moreover, static-based techniques are unable to detect obfuscated malware (malware with evasion behaviours) \cite{Zelinka_Amer_2019} \cite{HAN2019208}. Such obfuscated malware includes Agent Tesla, BitPaymer, Zeus Panda, and Ursnif, to name a few \cite{Obfuscat45:online}.

Dynamic or behaviour-based malware detection technique has addresses the above shortcoming observed in the signature-based techniques. In contrast to signature-based techniques, dynamic-based techniques can detect both obfuscated and zero-day malware attacks. They are implemented based on the dynamic malware analysis approach which allows monitoring the suspicious program's behaviours and vulnerability exploits by executing it in a virtual environment \cite{udayakumar2017dynamic}. Dynamic analysis can reveal behavioural features such as running processes, registry key changes, web browsing history (such as DNS queries), malicious IP addresses, loaded DLLs, API calls, and changes in the file system \cite{Han2019a} \cite{maniriho2021study}. Hence, the dynamic analysis produces the best representation of  malware behaviour since in many cases the behaviour of malware remain the same despite many variants it may have. As malware can use API calls to perform different malicious activities in a compromised system, tracing API calls reveals how a particular executable file behaves \cite{vemparala2019malware} \cite{Singh2020a}. Accordingly, Vemparala et al.'s work \cite{vemparala2019malware} demonstrated that the dynamic-based malware detection model outperforms the static model in many cases, after comparing their performance using extracted dynamic API calls and static sequence of opcodes features. This work is focused on analyzing dynamic-based API call sequences to identify malware.


Existing techniques for dynamic-based malware detection have used machine learning (ML) algorithms to successfully identify malware attacks. These algorithms learn from given data and make predictions on new data. over the last decade, the use of ML-based models has become more prevalent in the field of cybersecurity such as malware detection \cite{8405026} \cite{Gibert2020}.  Decision Trees (DT), Support Vector Machine (SVM), J48, K-nearest neighbour (KNN), Random Forest (RF), and Naïve Bayes are the most popular ML algorithms which are used to build malware detection \cite{Singh2020a} \cite{maniriho2021study} \cite{Gibert2020}. However, conventional machine learning techniques rely on manual feature extraction and selection process, which requires human expert domain knowledge to derive relevant or high-level patterns/features to be used to represent a set of malware and benign files. This process is known as manual feature engineering and is time-consuming and error-prone as it depends on a manual process, considering the current plethora of malware production. Deep learning (DL) algorithms have also emerged for malware detection \cite{Bostami2020} \cite{li2022novel}. Deep neural networks (DNNs) \cite{kim2018multimodal}, recurrent neural networks (RNNs) \cite{catak2020deep}, autoencoders \cite{Tirumala2020}, convolutional neural networks (CNNs) \cite{8802378} \cite{9204665}, and Deep Belief Networks (DBNs) \cite{8556824} are examples of DL algorithms that have been used in dynamic malware analysis \cite{Suaboot2020} \cite{amer2022malware}. Different from conventional ML techniques, DL algorithms can perform automatic feature extraction \cite{Tirumala2020}. 
 
On the other hand, the majority of existing ML and DL-based techniques operate as black boxes \cite{moraffah2020causal} \cite{mehrabi2019survey} \cite{maniriho2021study}. These models receive input $X$ which is processed through a series of complex operations to produce $Y$ as the predicted outcome/output. Nevertheless, these operations cannot be interpreted by humans, as they fail to provide human-friendly insights and explanations, for example, which features contributed to the final predicted outcome \cite{moraffah2020causal} \cite{mehrabi2019survey}. Therefore, by using explainable modules researchers and security analysts can derive more insight from the detection models and understand the logic behind the predictions \cite{ribeiro2016should}. The interpretation of the model’s outcome can help to assess the quality of the detection model in order to make the correct decision.

\subsection*{Motivation and Contributions}
\label{motiv-contrib}
Extracting behavioural features from malware executable files is a very critical task as malware can damage organizational resources such as corporate networks, confidential information, and other resources when they escape the analysis environment. Hence, obtaining a good and up-to-date behavioural representation of malware is a challenging task \cite{mimura2022applying}. In the case of API calls, having a good representation of API call features from benign and malware programs also remains a challenge as the number of API calls made by malware executable files is relatively long which makes their analysis and processing difficult \cite{Suaboot2020}. In addition, some of the previous techniques have used API calls to detect malicious executable files. Such API call-based malware detection techniques include the ones presented in \cite{Suaboot2020} \cite{Amer2020} \cite{Sihwail2019} \cite{Singh2020b} and \cite{amer2022malware}. Unfortunately, none of the previous work has attempted to reveal API call features that contributed to the final prediction/classification outcome. Practically, it is ideal to have a machine learning or deep learning model that can detect the presence of malicious files with high detection accuracy. However, the prediction of such a model should not be blindly trusted, but instead, it is important to have confidence about the features or attributes that contributed to the prediction.

In order to improve the performance of existing malware detection techniques, this work proposes MalDetConv, a new automated framework for detecting both known and zero-day malware attacks based on natural language processing (NLP) and deep learning techniques. The motivation for using deep learning is to automatically identify unique and high relevant patterns from raw and long sequences of API calls which distinguishes malware attacks from benign activities, while NLP techniques allow producing numerical encoding of API call sequences and capturing semantic relationships among them. More specifically, MalDetConv uses an encoder based on NLP techniques to construct numerical representations and embedding vectors (dense vectors) of each API call based on their semantic relationships. All generated embedding vectors are then fed to a CNN module to automatically learn and extract high-level features of the API calls. The final features generated by CNN are then passed to a bidirectional gated recurrent unit (BiGRU) feature learning module to capture more dependencies between features of API calls and produce more relevant features. We believe that combining CNN and BiGRU creates a hybrid automatic feature extractor that can effectively capture relevant features that can be used in detecting malicious executable files. Finally, features generated by the BiGRU module are fed to a fully connected neural network (FCNN) module for malware classification. 
We have also integrated LIME into our framework to make it explainable \cite{molnar2020interpretable}. LIME is a framework for interpreting machine learning and deep learning black box models proposed by Ribeiro et al. \cite{ribeiro2016should}. It allows MaldetConv to produce explainable predictions, which reveal feature importance, i.e., LIME produces features of API calls that contributed to the final prediction of a particular benign or malware executable file. Explainable results produced by LIME can help cybersecurity analysts or security practitioners to better understand MalDetConv’s predictions and to make decisions based on these predictions. The experimental evaluations conducted using different datasets show better performance of the MalDetConv framework over other state-of-the-art techniques based on API call sequences. Specifically, the following are the contributions of this work.

\begin{enumerate}
  \item This work contributes "MalBehavD-V1", a new behavioural dataset of API call sequences extracted from benign and malware executable files. We have used the MalBehavD-V1 dataset to evaluate the proposed framework and our dataset is made publicly available for use by the research community. Our dataset contains behaviours of 1285 new malware samples that appeared in the second quarter of 2021 and were not analyzed by the previous works. 
 \item We designed and implemented the MalDetConv framework using an NLP-based encoder for API calls and a hybrid automatic feature extractor based on deep learning techniques.
  \item MalDetConv uses LIME to provide explainable predictions which are essential for security analysts to identify interesting API call features which contributed to the prediction.
  \item Detailed experimental evaluations confirm the superiority of the proposed framework over existing  when detecting unseen malware attacks with an average of 97.48\% in the detection accuracy obtained using different datasets of API calls. Our framework outperforms existing techniques such as the ones proposed in \cite{Amer2020} \cite{karbab2019maldy} \cite{amer2022malware}.  
\end{enumerate}

\textbf{Structure:} The remainder of this paper is structured as follows. Section \ref{backgroud} presents the background and Section \ref{relaed-work} discusses the related works. Section \ref{proposed} presents the proposed method while Section \ref{result-discus} discusses the experimental results. Section \ref{limitations} presents limitations and future work. The conclusion of this work is provided in Section \ref{conclu}. 

\section{Background}
\label{backgroud}
This section discusses the prevalence of malware attacks in the Windows operating system (OS), Windows application programming interface (Win API), and malware detection using API calls. Moreover, it provides brief background on deep learning techniques, such as convolutional neural networks and recurrent neural networks.

\subsection{High Prevalence of Malware Attacks on Windows platforms}
\label{mal win_os}
\begin{figure*}[h]
  \centering
  \includegraphics[width=0.73\linewidth]{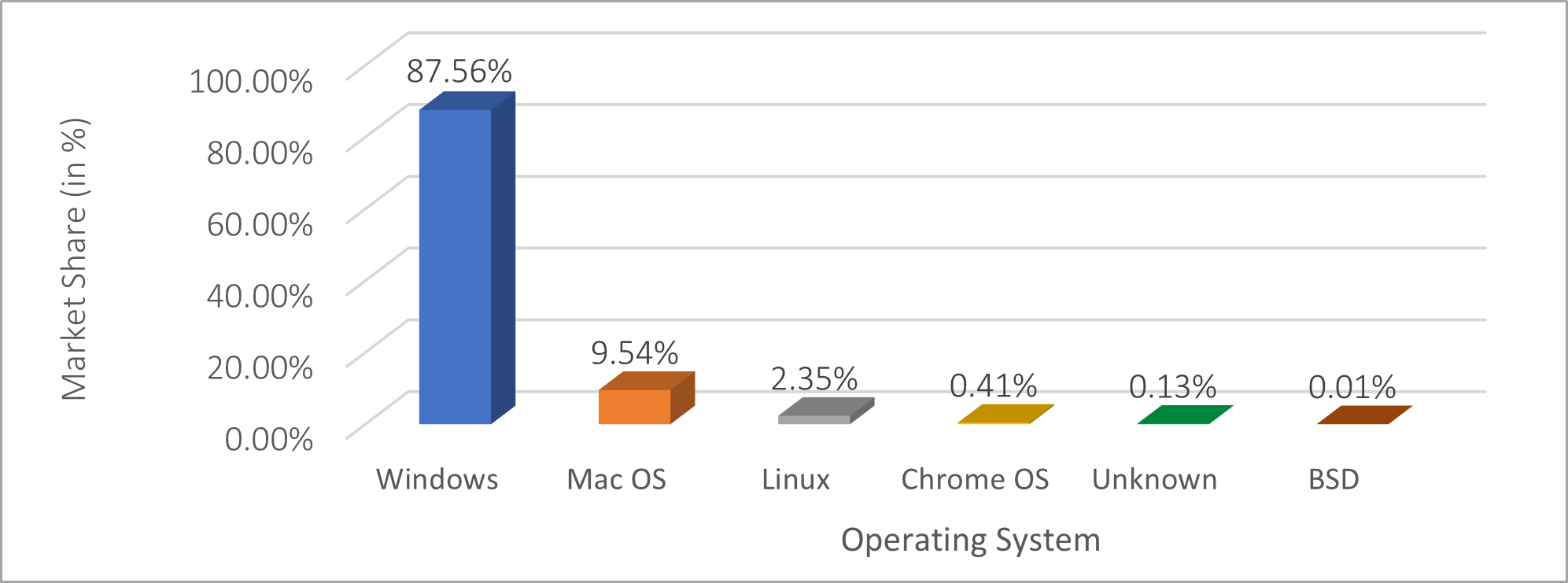}
  \caption{Distribution of global market share per each desktop operating system, source \cite{Operatin76:online}.}
  \label{os market share}
\end{figure*}

Malware production is increasing enormously with millions of threats targeting Windows OS. Developed by Microsoft, Windows is the most widely used and distributed desktop OS with the highest global market share \cite{Operatin76:online}. The distribution of this OS market share can also be viewed in Figure \ref{os market share}. Consequently, its popularity and widespread usage give many opportunities to cybercriminals to create various malware applications (malicious software) against Windows-based systems or devices \cite{Over100M74:online} \cite{WindowsU4:online}.


\subsection{Windows API}
\label{mal win_os}
The Windows application programming interface (API), also known as Win32 API, is a collection of all API functions that allow windows-based applications/programs to interact with the Microsoft Windows OS (Kernel) and hardware \cite{Quickint67:online} \cite{Silberschatz2018} \cite{Uppal2014}. Apart from some console programs, all windows-based applications must employ Windows APIs to request the operating system to perform certain tasks such as opening and closing a file, displaying a message on the screen, creating, writing content to files, and making changes in the registry. This implies that both system resources and hardware cannot be directly accessed by a program, but instead, programs need to accomplish their tasks via the Win32 API. All available API functions are defined in the dynamic link libraries (DLLs), i.e., in .dll files included in C:\textbackslash Windows\textbackslash System32\textbackslash *.
For example, many commonly used libraries include Kernel32.dl, User32.dll, Advapi32.dll, Gdi32.dll, Hal.dll, and Bootvid.dll \cite{Programm42:online}.

\subsection{API calls Monitoring}
\label{api monitoring}
Generally, any windows-based program performs its task by calling some API functions. This functionality makes Win32 API one of the important and core components of the Windows OS as well as an entry point for malware programs targeting the windows platform, since the API also allows malware programs to execute their malicious activities. Therefore, monitoring and analyzing Windows API call sequences gives the behavioural characteristics that can be used to represent benign and malware programs \cite{Ammar206656} \cite{ki2015novel}. API calls analysis reveals a considerable representation of how a given malware program behaves. Therefore, monitoring the program’s API call sequences is by far one of the most effective ways to observe if a particular executable program file has malicious or normal behaviours \cite{gupta2016malware} \cite{zhao2019feature} \cite{ALAEIYAN201976} \cite{DING201873} \cite{Amer2020}.


\subsection{Deep Learning Algorithms}
\label{deep-l}
Deep learning (DL) techniques are subsets of machine learning techniques that use artificial neural network architectures to learn and discover interesting patterns and features from data. DL network architectures can handle big datasets with high dimensions and perform automatic extraction of high-level abstract features without requiring human expertise in contrast to ML techniques \cite{SHARMA202124} \cite {YUAN2021102221}. DL algorithms are designed to learn from both labeled and unlabeled datasets and produce highly accurate results with low false-positive rates \cite{najafabadi2015deep}. By using a hierarchical learning process, deep learning algorithms can generate high-level complex patterns from raw input data and learn from them to produce intelligent classification model which performs classification tasks, making deep models valuable and effective for big data manipulation. The multi-layered structure adopted by deep learning algorithms gives them the ability to learn relevant data representations through which low-level features are captured by lower layers and high-level abstract features are extracted by higher layers \cite {rafique2020malware} \cite{PINHERO2021102247}. The next section introduces CNN and recurrent neural networks, which are some of the popular categories of DL algorithms. 

\subsubsection{Convolutional Neural Network}
\label{cnns}
\begin{figure*}[!ht]
  \centering
  \includegraphics[width=0.90\linewidth]{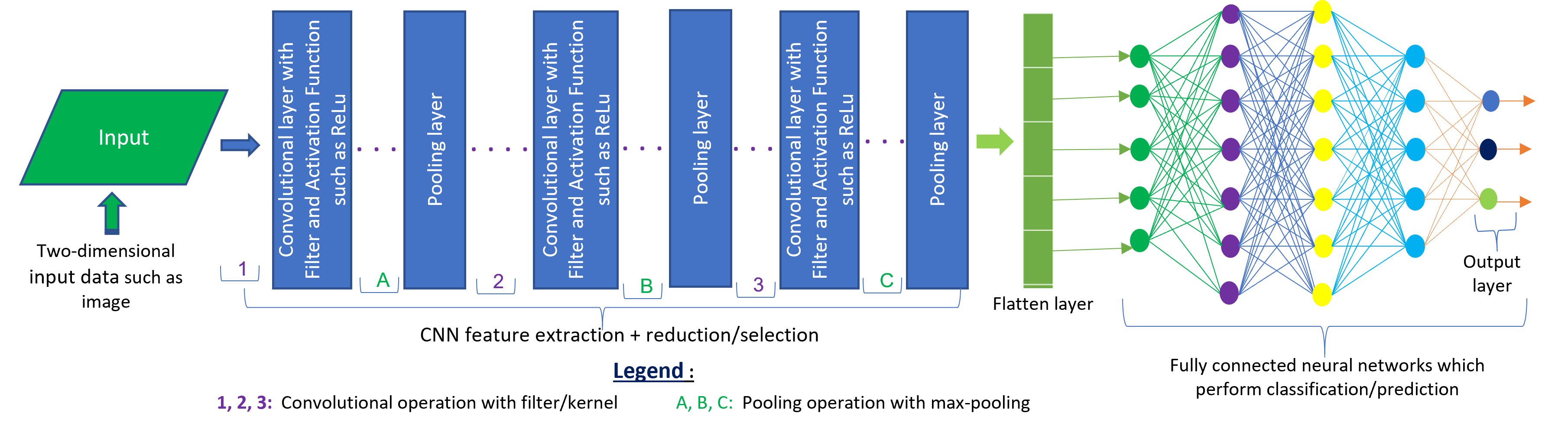}
  \caption{General architecture depicting how CNNs  work.}
  \label{2dcnn}
\end{figure*}

Convolutional neural network (CNN) is a category of deep learning techniques that gained popularity over the last decades.
Inspired by how the animal visual cortex is organized \cite{hubel1968receptive} \cite{fukushima1979neural}, CNN was mainly designed for processing data represented in grid patterns such as images. CNN  has been successfully used to solve computer vision problems \cite{khan2018guide} and has attracted significant interest across various image processing domains such as radiology \cite{lundervold2019overview}. Convolutional neural networks (CNNs) are developed to automatically learn and extract high-level feature representation from low patterns of raw datasets. The architecture of the CNN technique has three main components or layers that are considered its main building blocks (see Figure \ref{2dcnn}). These components include the convolutional layer, pooling layer, and fully connected layer (also known as the dense layer). The convolution layer and pooling layers are responsible for feature extraction and selection/reduction while the dense layer receives the extracted features as input, learns from them, and then performs classification which gives the output, e.g., a class of a given instance.

CNN differs from the existing conventional machine learning techniques as follows.
\begin{itemize}
    \item Most of the current traditional machine learning techniques are based on manual or hand-crafted feature extraction and selection techniques which are followed by the learning and classification stages performed by the machine learning classier \cite{Singh2020b}.
   \item In contrast, CNN network architectures do not use manual crafted-feature extraction and selection as they can automatically perform both operations and produce high-level feature representations through convolution and pooling operations \cite{lundervold2019overview}.
   \item CNN uses several learnable parameters to effectively learn and select relevant features from input data \cite{lundervold2019overview} \cite{khan2018guide}.
\end{itemize}

CNN can have single or multiple convolution and pooling layers, with the convolution layer being its core component. The goal of the convolutional layer is to learn different high-level feature representations from the input data using several learnable parameters called filters or kernels which generate different feature maps through convolution operations. During convolution, the defined filters are convolved to a window of the same size from input data which extracts features from different positions of the input. The element-wise multiplication is performed between the filter and each input window size.  Strides and padding are also important parameters of the convolutional layer \cite{o2015introduction}. Stride is the defined step to which the filter should be moved over the input (horizontally or vertically) for 2-D data.  In the case of 1-D input data, the filter moves in one direction (horizontally) with stride. Padding allows producing feature maps with a size similar to the input data and helps to protect the boundary information from being lost as the filter moves over the input data, with zero-padding being the most used method. The convolved results are passed to a non-linear activation function before being fed to the next layer.

The pooling layer (also known as the sub/down-sampling layer) selects a small region of each output feature map from the convolution and sub-samples them to generate a single reduced output feature map. Max-overtime pooling, average/mean pooling, and min pooling are different examples of pooling techniques. Note that the pooling operation reduces the dimension of input data and results in a few parameters to be computed in the upper layers, making the network less complex while keeping relevant feature representation at the same time \cite{yamashita2018convolutional}.

The next layer (considered the last layer of CNN) is a fully connected neural network (FCNN) layer that receives the pooling layer’s output as input and performs classification/predictions. It is an artificial neural network with hidden layers and an output layer that learn and perform classification through forward propagation and backpropagation operations. Hence, a typical CNN model can be viewed as an integration of two main blocks, namely, feature extraction and selection block, and classification block. One dimensional CNN (1-D CNN) is a version of 2-D CNNs and has been recently presented in various studies \cite{kiranyaz2015real} \cite{kim2014convolutional} \cite{fesseha2021text} \cite{avci2018efficiency}. All these studies have proven that 1-D CNNs are preferable for certain applications and are advantageous over their 2-D counterparts when handling 1-D signals. One dimensional (1D) represents data with one dimension such as times series data and sequential data. Another recent study has mentioned that 1-D CNNs are less computationally expensive compared to 2-D CNNs and do not require graphics processing units (GPUs) as they can be implemented on a standard computer with a CPU and are much faster than 2-D CNNs \cite{kiranyaz20211d}. 1-D CNN architectures have been also successful in modeling various tasks such as solving natural language processing (NLP) problems. For example, in the previous work, CNNs were successfully applied to perform text/documents classification \cite{kim2014convolutional} \cite{WANG2016806} \cite{johnson2015effective} and sentiments classification \cite{LIU202021}.

\subsubsection{Recurrent neural networks}
\label{rnns}

\begin{figure*}[!ht]
  \centering
  \includegraphics[width=0.93\linewidth]{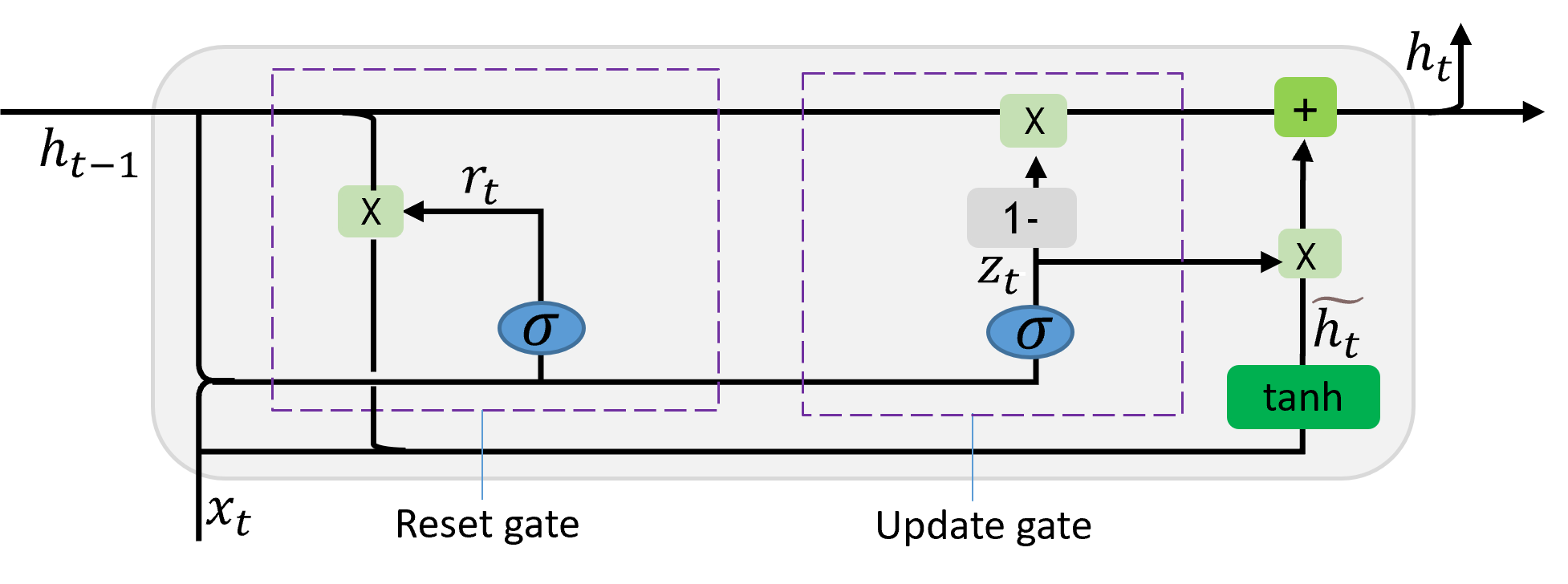}
  \caption{The typical network architecture of the gated recurrent unit algorithm.}
  \label{GRU-arch}
\end{figure*}

A Recurrent neural network (RNN) is a type of DL network architecture which is suitable for modeling sequential data. RNN uses a memory function that allows them to discover important patterns from data. However, traditional/classic RNNs suffer from vanishing gradients (also known as gradient explosion) and are unable to process long sequences \cite{lynn2019deep}. To address this problem, Hochreiter et al. \cite{hochreiter1997long} proposed Long short-term memory
 (LSTM), an improved RNN algorithm that performs well on long sequences. The Gated Recurrent Unit (GRU) was later implemented by Chao et al. \cite{cho2014learning} based on the LSTM. As depicted in Figure \ref{GRU-arch}, a GRU network uses a reset gate and update gate to decide which information needs to be passed to the output. A reset gate is used to determine which information to forget in the hidden state of the previous time step/moment, where the information in the previous time step will be forgotten when the value of the reset gate is close to 0, otherwise it will be retained if the value is close to 1. The update gate decides how much information needs to be passed to the current hidden state. GRU operations are computed using mathematical equations in \eqref{updateget}, \eqref{restgate},\eqref{info-retained}, and \eqref{nxt-state}. 

\begin{equation}
\label{updateget}
z_{t}=\sigma(w_{zx}x _{t }+u_ {zh} h_{t-1}+b_{z})
\end{equation}
\begin{equation}
\label{restgate}
r_{t}=\sigma(w_{rx}x_{t}+u_{rh}h_{t-1}+b_{r})
\end{equation}
\begin{equation}
\label{info-retained}
\tilde{h}_{t}= tanh(w_{hx}x_{t}+r_{t}\odot u_{hh}{h_{t-1}}+b_{h})
\end{equation}
\begin{equation}
\label{nxt-state}
h_{t} = (1-z_{t})\odot \tilde{h}_{t}+z_{t} \odot h_{t-1})
\end{equation}

In Figure \ref{GRU-arch}, $z_{t}$, $r_{t}$, $\tilde{h}_{t}$ and $h_{t}$ denote the update gate, reset gate, candidate hidden state of the currently hidden node and current hidden state, respectively. The symbol $\sigma$ denotes the Sigmoid activation, $x _{t}$ is the current input, $tanh$ is the  hyperbolic tangent activation function, $w$ and $u$ are weights matrices to be learned,  $\odot$ is the Hadamard product of the matrix, while $b_{z}, b_{r},$ and $b_{h}$ denotes the bias. The values of $z_{t}$ and $r_{t}$ are between 0 and 1 and when modeling sequences the reset gate operates on $h_{t-1}$ to record all necessary information in the memory content. The information to be forgotten in the current memory content is determined by the reset gate after obtaining the Hadamard product \cite{horn2012matrix}. All important information is recorded by $\tilde{h}_{t}$ through the input information and reset gate. The update gate acts on ${h_{t-1}}$ and $\tilde{h}_{t}$ and forwards it to the next unit. As it can be seen from equation \ref{nxt-state},  the expression $(1-z_{t})$ decides which information needs to be forgotten and then the associated information in the memory content is updated. Hence, the required information to be retained is decided by $ht$  at $\tilde{h}_{t}$ and ${h_{t-1}}$ via the update gate. Every hidden layer of the GRU network has a reset and update gate which are separated, and based on the current input information and the information from the previous time-step, layers produce different dependent relationships among features. GRU uses a simple network architecture and has shown better performance over regular LSTMs \cite{chung2014empirical}. Bidirectional GRU (BiGRU) is a variant of GRU that models information in two directions (right and left direction) \cite{vukotic2016step}. Accordingly, studies have shown better performance with reversed sequences, making it ideal to model sequences in both directions in some cases \cite{vukotic2016step}. 

\section{Related Works}
\label{relaed-work}
This section presents related works on static and behaviour-based malware detection. However, all works that use images of  binary files and hybrid malware detection techniques that use a combination of static and behavioural features are beyond the scope of this literature. In addition, works that do not use windows executable files for experimental analysis are also not included in this section.

\subsection{Static and Dynamic-based malware detection techniques}
\label{Static and Dynamic-based}
There have been significant efforts in the recent works on malware detection through static and behaviour-based analysis. Static API calls extracted after disassembling benign and malware executables with IDA Pro Disassembler were mined using a hybrid of SVM wrapper and filter methods \cite{huda2016hybrids}. Different subsets of features selected by the SVM wrapper were combined with feature subsets generated by the filter method to construct a good representation of API call features and the final feature set was then employed to build different hybrid SVM-based malware detection models. While the model performs well when classifying malware based on API call features, there is no clear approach for selecting filters, which can affect the overall performance, especially for the hybrid methods. The work presented by Naik \cite{naik2021fuzzy} has implemented a fuzzy-import hashing approach to measure similarity between malware and benign files and their results were further compared with YARA rules. The dataset with four ransomware families namely, WannaCry, Locky, Cerber Ransomware, and CryptoWall was used to evaluate the performance of their proposed approach. YARA rules were generated from strings extracted from these ransomware samples and different Fuzzy-Import Hashing approaches (SDHASH, IMPHASH, SSDEEP, etc.,) were tested.  The results reveal that there is a high similarity between ransomware from the same family and a high dissimilarity between ransomware from different families. API calls  and API calls' statistics extracted from binary files through the static malware analysis were used in the detection model proposed in Huda et al.'s work \cite{huda2017fast}. The step-wise binary logistic regression (SBLR) model was applied to select a subset of specific features and thereafter, the Decision Tree and SVM algorithms were trained and tested to detect malicious activities. While the focus of their work was to reduce the computation complexity, this technique depends on a linear approach which can affect the overall detection accuracy of the proposed model. An instruction sequence-based malware detection framework was proposed in Fan et al.’s work \cite{fan2016malicious} where sequences were extracted from executable samples and malicious sequential patterns were discovered using a sequence mining approach that works in combination with All-Nearest Neighbour (ANN) classifier. Based on the experimental results, this method can identify malicious patterns from given suspicious executable programs with improved detection accuracy.

A signature-based malware detection approach was implemented using deep autoencoders in \cite{Tirumala2020}. An opcode sequence analysis method was used to construct a static-based method that effectively detects malware attacks \cite{sun2019opcode}. Opcode sequences were statically extracted from 32-bit and 64-bit Windows EXE files and the occurrence of each opcode  was computed using the term frequency-inverse document frequency (TF–IDF) to obtain a feature set that was used to train KNN, Decision Tree, Adaboost, RF, Bagging and backpropagation neural networks detection models. A total of 20,000 files (malware and benign) was used in the experiment and k-fold cross-validation was used to evaluate the performance of the proposed methods. The results reveal that their malware detection systems can identify and classify different malware attacks with the better performance achieved by the Adaboost model. Logistic regression was used to learn relevant features (to perform domain knowledge operation) from raw bytes to reveal the best byte n-grams features which were further employed to train the Random Forest and Extra Random Trees using the JSAT library \cite{raff2017learning}. Furthermore, they have also built a long short-term memory (LSTM) and fully connected neural networks detection approach using raw bytes and Keras library \cite{Gettings10:online}.  Their performance evaluation shows that neural network models trained on raw bytes features without performing explicit feature generation can outperform the performance of domain knowledge methods that fully depends on constructing explicit features from raw bytes extracted from PE headers \cite{raff2017learning}. D’Onghia et al. \cite{d2022apicula} proposed Apícula, a static analysis-based tool that uses the Jaccard index to identify malicious  API calls presented in bytes streams such as network traffic, object code files, and memory dumps. The work presented by Kundu et al. \cite{kundu2021empirical} has improved the performance of a Light Gradient Boosted Machine (LightGBM) approach for detecting and classifying malicious software using two datasets, the Microsoft EMBER dataset and another private dataset collected from an anti-virus company. The optimization (hyper-parameter tuning) was performed using Microsoft Neural Network Intelligence (NNI) and AutoGluon-Tabular (AutoGluon) automated machine learning frameworks.  The AutoGluon was released by Amazon and has been implemented with various ML algorithms including KNN, LightGBM, and Multi-Layer Perceptron (MLP). The results obtained after performing different empirical analyses demonstrate that there is an improvement in tuned models compared to the baseline models. Kale et al.’s \cite{kale2022malware} used opcode sequences from EXE files to build malware detection approaches based on RF, SVM, KNN, and CNN where feature extraction and representation were performed using different techniques such as Word2Vec, BERT, HMM2Vec, and ELMo. Yeboah et al.\cite{yeboah2022nlp} proposed a CNN-based approach that detects malicious files using operational code (opcode). 

Nevertheless, cybercriminals can easily modify the executable binary file’s code and disguise a small piece of encrypted malicious code, which can be decrypted during its execution/runtime \cite{safebreach2017}. Such malicious code can even be downloaded remotely and get inserted into the program while running. Accordingly, the work presented by \cite {mar4413009} has also revealed that it is possible for encrypted and compressed code to be decrypted and unpacked when loaded into memory during runtime. Additionally, some sophisticated malware uses more than two stages to attack the victim’s systems. For instance, malware designed to target specific platforms such as Windows OS will first check if it is running in the targeted OS. Once detected, a malicious payload can automatically be downloaded from the cybercriminal's server and get loaded into memory to perform malicious tasks. Malware with the ability to evade antivirus and utilizes it to transfer users’ stolen confidential information from the comprised users without any notice was reported in \cite {safebreach2017}. Recently advanced malware variants use varieties of sophisticated evasion techniques including obfuscation, dead code insertion, and packing \cite{Yucel2020} \cite{ye2017m} to evade static-based malware detection, making them more vulnerable and unable to detect new sophisticated malware attacks. Fortunately, the dynamic malware analysis approach  can handle obfuscated malware and  consequently, many of the recent works were focused on dynamics-based approaches/techniques. The work proposed by Vemparala et al. \cite{vemparala2019malware} extracted API call sequences through dynamic analysis and employed them to train and test both hidden Markov Models and Profile Hidden Markov Models to observe the efficiency of the dynamic analysis approach while detecting malware. Their study has used a dataset of seven malware types with some malware programs such as Zbot and Harebot that have the capability of stealing confidential information. The results from their experiment show that the HMM-based malware detection approach using behavioural/dynamic sequences of API calls outdoes both static and hybrid-based detection techniques. Nevertheless, the authors have used an outdated dataset, which can prevent the model from identifying new sophisticated malware attacks, considering the rapid scale of different malware variants.

The work in \cite{ki2015novel} has adopted the DNA sequence alignment approach to design a dynamic analysis-based method for malware detection. Common API call features were dynamically extracted from different categories of malware.  Their experimental outcome has revealed that some of the malicious files possess common behaviours/functions despite their categories, which may be different. In addition, their study has also indicated that unknown or new malware can be detected by identifying and matching the presence of certain API calls or function calls as malware programs perform malicious activities using almost similar API calls. However, the limitation of DNA sequence approaches is that they are prone to consuming many resources and require high execution time, making them computationally expensive, considering the high volume of the emerging malware datasets. 
Seven dynamic features, namely, API calls, mutexes, file manipulations/operations, changes in the registry, network operations/activities, dropped files, and printable string information (PSI) were extracted from 8422 benign and 16489 malware using the Cuckoo sandbox \cite{Singh2020a}. PSI features were processed using a count-based vector model (count vectorization approach) to generate a feature matrix representing each file and thereafter, truncated singular value decomposition was utilized to reduce the dimension of each generated matrix. Furthermore, they have computed Shannon entropy over PSI and API call features to determine their randomness. Using PSI features, their model has achieved the detection accuracy of 99.54\% with the Adaboost ensemble model while the accuracy of 97.46\% was yielded with the Random Forest machine learning classifier.  However, while their method shows an improvement in the accuracy of the detection model using PSI features, it is worth mentioning that the count vectorization model applied while processing features does not preserve semantic relationship/linguistic similarity between features or word patterns  \cite{CountVe93:online} \cite{mandelbaum2016word}. The count vectorization model is unable to identify the most relevant or less relevant features for the analysis, i.e., only words/features with a high frequency of occurrence in a given corpus are considered as the most statistically relevant words.

Cuckoo sandbox was used to generate behavioural features of benign and malware files during their execution time in the work carried out in \cite{Pirscoveanu2015}. They have used Windows API calls to implement a Random Forest-based malware detection and classification approach that achieved the detection accuracy of 98\%. Malicious features were extracted from about 80000 malware files including four malware categories, namely, Trojans, rootkit, adware, and potentially unwanted programs (PUPs) downloaded from VirusTotal and VirusShare malware repository. The detection model was tested on 42000 samples and the results show an improvement in the detection of malware attacks. However, looking at the detection outcome, this approach only performs well when detecting Trojans. For instance, it achieved the true positive rate (TPR) of 0.959 when detecting and classifying Trojans while the TPR of 0.777, 0.858, and 0.791 were achieved while detecting rootkit, adware, and PUPs, respectively.  In the work presented by Suaboot et al. \cite{Suaboot2020} a subset of relevant features was extracted from API call sequences using a sub-curve Hidden Markov Model (HMM) feature extractor. Benign and malware samples were executed in the Cuckoo sandbox to monitor their activities while executing in a clean isolated environment residing in Windows 7-32 bits machine. The API call features selected from the behavioural reports of each executable program file (benign and malware) were then used to train and test a behaviour-based malware detection model. Only six malware families (Keyloggers, Zeus, Rammit, Lokibot, Ransom, and Hivecoin) with data exfiltration behaviours were used for the experimental evaluation. Different machine learning algorithms such as Random Forest (RF), J48, and SVM were evaluated, and the RF classifier achieved better prediction with an accuracy average of 96.86\% compared to other algorithms. Nevertheless, their method was only limited to evaluating executable program files, which exfiltrates confidential information from the compromised systems. In addition, with 42 benign and 756 malware executable programs which were used in the experimental analysis, there is a significant class imbalance in their dataset which could lead to the model failing to predict and identify samples from minority classes despite its good performance.

A backpropagation neural networks (BPNNs) approach for malware detection was proposed in \cite{Pan2016}.  The proposed BPNNs–based approach learns from features extracted from benign and malware behavioural reports to classify malware attacks. A dataset of 13600 malware executable files was collected from Kafan Forum and  the dataset has ten malware families which were labeled using Virscan. They also have used the HABO online behaviour-based system, which captures the behaviours of each file and generates a behavioural report having all activities performed by the executable file. Behavioural features such as mutexes, modified registry keys, and created processes, to name a few, were used to train and test the proposed BPNNs approach to perform binary classification, which led to the accuracy of 99\%. Note the BPNN was built using a sub-behavioural feature generated based on a count-based method which is limited to using an occurrence matrix that does not provide any relationship between the selected sub-behavioural feature set.  Malware attacks can be detected using the detection techniques implemented based on the longest common substring and longest common subsequence methods suggested in \cite{mira2016novel}. Both methods were trained on behavioural API calls, which were captured from 1500 benign and 4256 malware files during dynamic analysis. A Malware detection approach based on API calls monitored by the Cuckoo sandbox was designed in the work proposed by Kyeom et al. \cite{cho2016malware}. The analysis was carried out using 150 malware belonging to ten malware variants and the detection was achieved by computing similarity between files based on the extracted sequence of API calls features using the proposed sequence alignment approach. Their results also show that similar behaviours of malware families can be found by identifying a common list of invoked API call sequences generated during the execution of  executable program. Unfortunately, this method cannot suitable for high-speed malware attacks detection as it fully relies on the pairwise sequence alignment approach, which introduces overheads. The work in \cite{stiborek2018multiple} has proposed multiple instance-based learning (MIL) approach that exploits the performance of several machine learning classifiers (SVM, KNN, RF, etc.) to detect malicious binary files. The MIL was trained using different dynamic features such as network communication (operations), the structure of file paths, registry, mutexes, keys, and error messages triggered by the operating system. Behaviours of benign and malicious files were monitored in the sandbox and a similarity-based method was used in combination with clustering which allows similar systems resources to be grouped together. MIL has achieved a detection accuracy of 95.6\% while detecting unknown malware files.

Several graph-based techniques for malware detection were proposed in the previous studies \cite{wucher7867799} \cite{6703684} \cite{6679854} \cite{9318301} \cite{ding2018malware} and \cite{pei2020amalnet}, to mention a few. All of these approaches have used graphs to represent behaviours of executable files and the detection models learned from the generated graphs. Nevertheless, the complexity of graph matching is one of the major issues with graph-based techniques, i.e., as the graph’s size increases the matching complexity, the detection accuracy of a given detection model decreases \cite{Singh2020a}. Grouping a common sequence of API calls in a single node is often applied to decrease the matching complexity of graphs. Although this approach does not yield better detection accuracy, it is harder for a cyber attacker to modify the behaviours of a malware detection model based on the graph method \cite{9318301}. NLP techniques were applied to analyze/mine the sequence of API calls in the malware detection approach implemented in \cite{8233569}. The API calls were processed using the n-gram method and the weights were assigned to each API call feature using the term frequency-inverse document frequency (TF-IDF) model. The main goal of the TF-IDF is to map n-grams of API calls into numerical features that are used as input to machine learning or deep learning algorithms. Given an input of behavioural report extracted from malware and benign files in a dynamic isolated environment, the TF-ID, processes the report to generate feature vectors by taking into account the relative frequency of available n-grams in the individual behavarioural report compared to the total number of the report in the dataset. The work in \cite{karbab2019maldy} has also relied on TF-IDF to compute input features for machine learning algorithms (CART, ETrees KNN, RF, SVM, XGBoost, and Ensemble). Another previous work in \cite{catak2020deep} has used LSTM and TF-IDF model to build a behaviour-based malware detection using API call sequences extracted with cuckoo sandbox in a dynamic analysis environment.

The work in \cite{mimura2022applying} has  used the term frequency-inverse document frequency to process printable strings extracted from malware executable files after dynamic analysis. TF-IDF and Anti-colony optimization (Swarm algorithm) were used to implement a behavioural graph-based malware detection method based on dynamic features of API calls extracted from executable \cite{amer2022malware}. Unfortunately, like the count vectorization model, the TF-IDF text processing model does not reveal or preserve the semantic relationship/similarity that exists between words. In the case of malware detection, this would be the similarity between API calls or another text-based feature such as file name, dropped messages, network operations such as contacted hostnames, web browsing history, and error message generated while executing the executable program file. Liu and Wang have \cite{liu2019robust} used Bidirectional LSTM (BLSTM) to Build an API call-based approach that classifies malware attacks with an accuracy of 97.85\%. ALL sequences of API calls were extracted from 21,378 and were processed using the word2vec model. In Li et al.  \cite{li2022intelligent} a graph convolutional network (GCN) model for malware classification was built using sequences of API calls.  Features were extracted using principal component analysis (PCA) and Markov Chain. A fuzzy similarity algorithm was employed by Lajevardi \cite{lajevardi2022markhor} to develop dynamic-based malware detection techniques based on API calls. Maniath et al. \cite{8358312} proposed an LSTM-based model that identifies ransomware attacks based on Behavioural API calls from Windows EXE files generated through dynamic analysis in the Cuckoo sandbox. Chen et al.’s work \cite{chen2022cruparamer} proposed different malware detection techniques based on CNN, LSTM, and bidirectional LSTM models. These models were trained on raw API sequences and parameters traced during the execution of malware and benign executable files. An ensemble of ML algorithms for malware classification based on  API call sequences was recently implemented in the work presented in  \cite{9777099}. Convolutional neural networks and BiLSTM were used to develop a malware classification framework based on sequences of API calls extracted from executables files \cite{li2022novel}. Two detests were used by Dhanya et al. \cite{dhanya2022performance} to evaluate the performance of various machine learning-based ensemble models (AdaBoost, Random Forest, Gradient descent boosting, XGBoost, Stacking, and Light GBM) for malware detection. Their study has also evaluated  DL algorithms such as GRU, Graph Attention Network, Graph Convolutional Network, and LSTM. The work proposed in \cite{abbasi2022behavior} has employed a dataset of API invocations, Registry keys, files/directory operations, dropped files, and embedded strings features extracted using the Cuckoo sandbox to implement a particle swarm-based approach that classifies ransomware attacks. Jing et al. \cite{jing2022ensemble} proposed Ensila, an ensemble of RNN, LSTM, and GRU for malware detection which was trained and evaluated on dynamic features of API calls.  

\subsection{API calls Datasets}
\label{api-calls data}
Some of the previous studies presented above have generated datasets of API calls and have made them public for the research community focusing on malware detection. Examples of such datasets include the ones presented in \cite{PDFWindo91:online}, \cite{ceschin2018need}, \cite{ki2015novel} and \cite{GitHuble77:online}. Nevertheless, many existing studies did not share their datasets and as result, a few datasets are publicly available for the research community. The existing datasets are also outdated as they are not regularly updated to include new behavioural characteristics of malware. This situation hinders the development and evaluation of new malware detection models as building these models requires updated datasets having the necessary behavioural characteristics of new malware variants. Therefore, to contribute to the existing challenge related to the availability of benchmark datasets \cite{mimura2022applying}, this work generates a new behavioural dataset of API calls based on Windows executable files of malware and benign using the Cuckoo sandbox. 

More specifically, we generate MalbehavD-V1, a new dataset that has the behavioural characteristics of current emerging malware such as ransomware, worms, Viruses, Spyware, backdoor, adware, keyloggers, and Trojans which appeared in the second quarter of 2021. The dataset has been processed to remove all inconsistencies/noise, making it ready to be used for evaluating the performance of deep learning models. The dataset is labeled and the hash value for each file has been included to avoid duplication of files while extending the dataset in the future, which makes it easier to include behavioural characteristics of newly discovered malware variants in the dataset or combine the dataset with any of the existing datasets of API calls extracted from Windows PE files through dynamic analysis. More details on the dataset generation are presented in Section  \ref{proposed} and the dataset can be accessed from the GitHub repository in \cite{mpascoMa95:online}.  


\section{Proposed MalDetConv Framework}   
\label{proposed}
\begin{figure*}[!ht]
  \centering
  \includegraphics[width=0.90\linewidth]{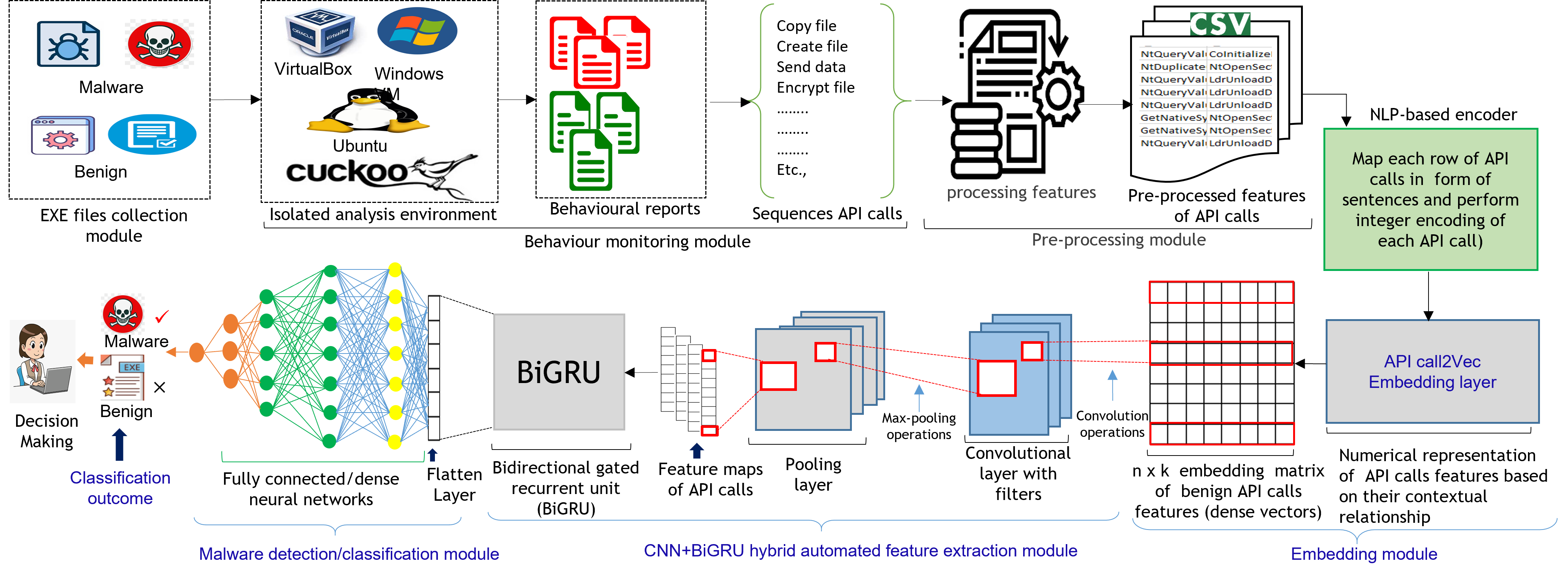}
  \caption{The proposed MalDetConv framework for behaviour-based malware detection.}
  \label{maldetconv}
\end{figure*}

The development of the proposed framework is mainly based on the dynamic malware analysis using natural language processing, convolutional neural networks, and the bidirectional gated recurrent unit. The benign and malware program’s dynamic/behavioural reports are created by analysing sequences of API calls in the Cuckoo sandbox. Given a program’s API call sequences represented in form of text, the proposed method uses an encoder based on the word embedding model to build dense vector representations of each API call which are further fed to a CNN-BiGRU automatic hybrid feature extractor. The architecture of the proposed framework is depicted in Figure \ref{maldetconv} and consists of six main components/modules, namely, executable files collection module, behaviour monitoring module, pre-processing module, embedding module, hybrid automatic feature extraction module, and classification module. In the next section, we present the function of each module and how all modules inter-operate to detect malicious EXE files.

\subsection{Executable File Collection Module}
\label{mal and bin samples}
It is often challenging to find an up-to-date dataset of API calls of Windows executable files. For this reason, we have generated a new dataset of API calls which is used for the experimental evaluations. Different sources of malware executable files such as Malheur \cite{rieck2011automatic}, Kafan Forum \cite{Pan2016}, Danny Quist \cite{Sethi2017}, Vxheaven \cite{huda2016hybrids}, MEDUSA \cite{nair2010medusa}, and Malicia \cite{nappa2015malicia} were used in the previous studies. Unfortunately, these repositories are not regularly updated to include new malware samples. Hence, we have collected malware executable samples from VirusTotal \cite{VirusTot34:online}, the most updated and the world's largest malware samples repository. Nevertheless, considering millions of malware samples available in the repository, processing all malware samples is beyond the scope of this study. Thus, only malware samples submitted in the second quarter of 2021 were collected. We were given access to a Google drive folder having the malicious EXE files which are shared by VirusTotal. Benign samples were collected from CNET site\cite{FreeSoft10:online}. The VirusTotal online engine was used to scan each benign EXE file to ensure that all benign samples are clean. A total number of 2800 EXE files were collected to be analyzed in an isolated analysis environment using the dynamic analysis approach. However, we experienced issues while executing some files, resulting in a dataset of 2570 files (1285 benign and 1285 malware) that were successfully executed and analyzed to generate our MalbehavD-V1 benchmark dataset. Some benign files were excluded as they were detected as malicious by some of the anti-malware programs in the VirusTotal online engine \cite{VirusTot60:online} while some malware files did not run due to compatibility issues. The next sub-Sections (\ref{behmoni-module} and \ref{processing-module}) present the steps followed to generate the dataset.


\subsection {Behaviour Monitoring Module}
\label{behmoni-module}
\begin{figure*}[!ht]
  \centering
  \includegraphics[width=0.83\linewidth]{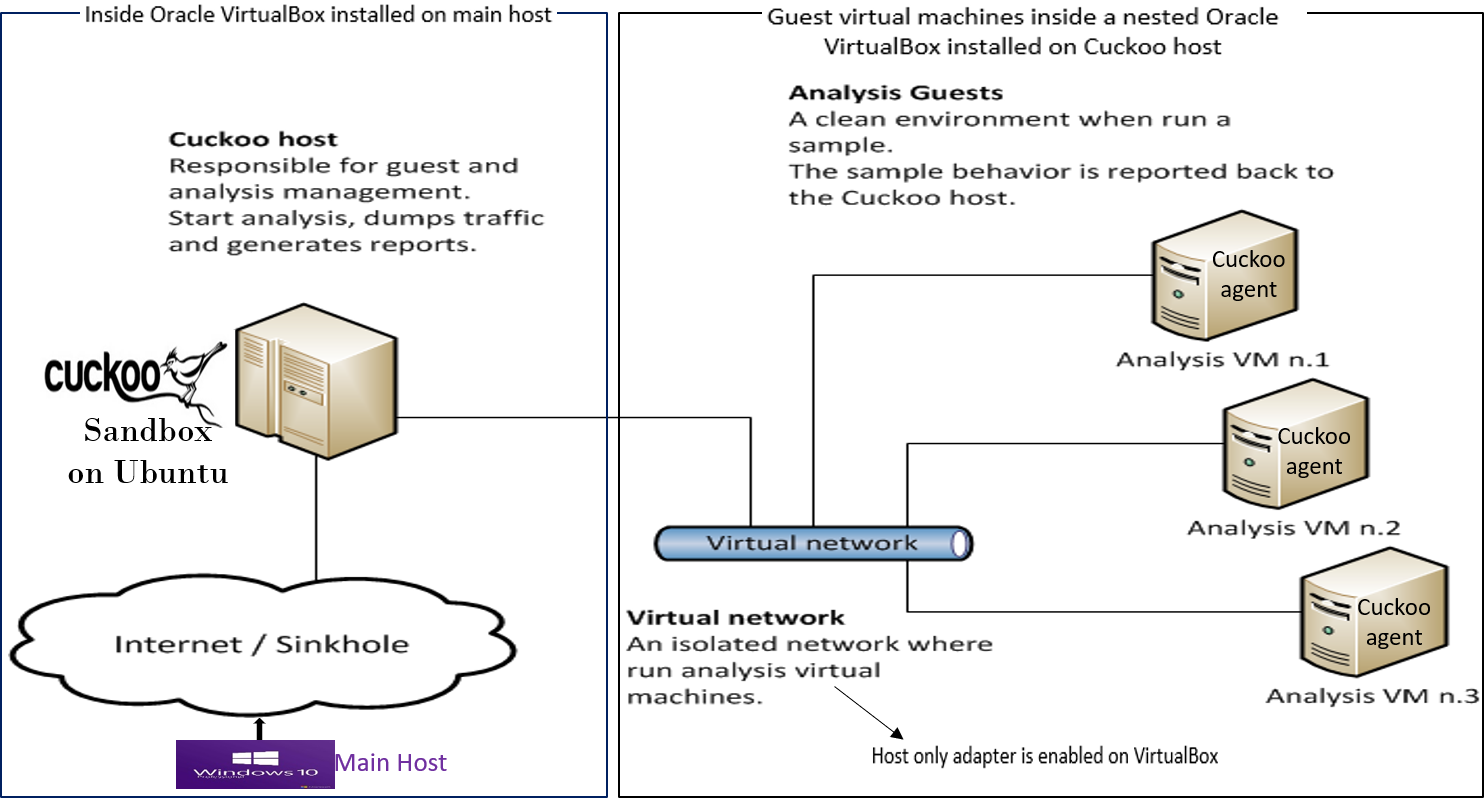}
 \caption{The architecture of our isolated dynamic malware analysis environment.}
  \label{analysisenv}
\end{figure*}

Monitoring and capturing sequences of API calls from benign and malware executable files through dynamic analysis mostly rely on the use of a sandbox environment as an analysis testbed. This is because automated sandboxes can monitor behaviours of malware executable programs during runtime while preventing them from damaging the host system or production environment. 
As shown in Figure \ref{analysisenv}, our isolated environment consists of the main host, host machine, Cuckoo sandbox, virtualization software, and virtual machines (VMs). The main host is a Windows 10 Enterprise edition (64bit) with direct access to the Internet. Considering that some very advanced malware can escape the sandbox analysis environment which could lead to serious damage and a disastrous situation when malware infects the production systems, we have used Oracle VirtualBox, one of the leading virtualization software to isolate the main host and the analysis environment which fully resides in Ubuntu 20.04.3 LTS (Focal Fossa) host machine. More specifically, Oracle VirtualBox 6.1 was installed on Windows 10 main host machine, and then Ubuntu host machine was installed inside the VirtualBox environment.

The Cuckoo sandbox \cite{CuckooSa69:online} and its dependency software including required analysis modules were installed on the Ubuntu machine. We have installed another VirtualBox that manages all Windows 7 Professional 64-bit SP1 analysis VMs on the Ubuntu machine.  A set of software was installed on each virtual machine and some security settings were disabled to make them more vulnerable to malware attacks. Examples of such software include Adobe reader 9.0, Java runtime environment (JRE 7), .NET Framework 4.0, Python 2.7 (32 bit), Pillow 2.5.3 (Python Imaging Library, 32 bit), Microsoft Office, and Cuckoo sandbox guest agent (32 bit). Python allows the Cuckoo guest agent to run while the pillow provides imaging functions such as taking screenshots. The cuckoo agent resides in each Windows VM under the Startup sub-directory. This allows the agent to automatically start whenever the VM boots up. A shared folder was configured in the Windows virtual machine to get all the above files, however, it was disabled after installing each file as some advanced malware looks for the shared folder to discover if they are running in a virtual environment or sandbox. As the Cuckoo guest agent must monitor each file’s behaviours while running and send all captured data back to the host, an Internet connection is needed between the Windows analysis VMs and Ubuntu host machine. Thus, we have set the network adaptor to "host-only Adaptor" in the VirtualBox to allow the network traffic to only be routed between the Windows virtual machines and Ubuntu host machine without reaching the main host (Windows 10 Enterprise host machine).

\begin{figure*}[!ht]
  \centering
  \includegraphics[width=0.75\linewidth]{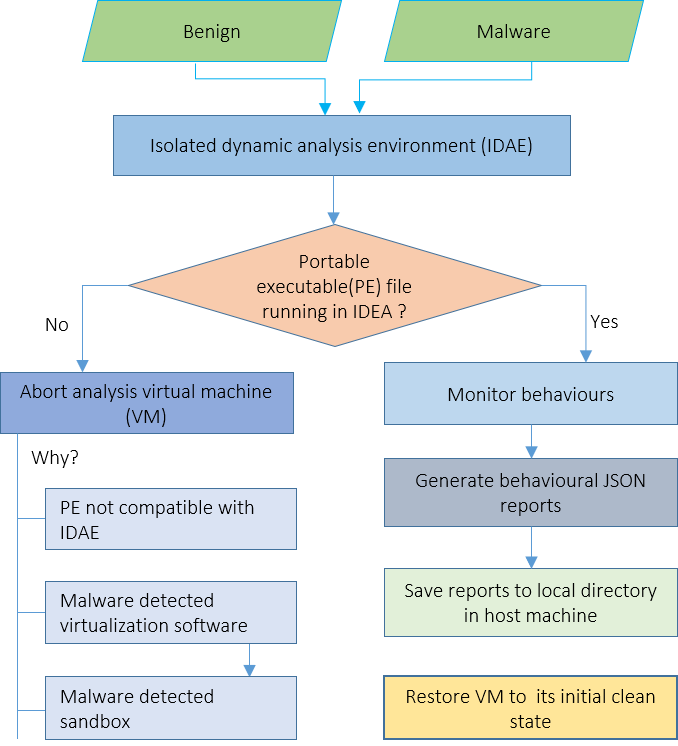}
  \caption{The workflow for submitting and executing each EXE file in the Cuckoo sandbox.}
   \label{submitmal}
\end{figure*}

\begin{figure*}[!ht]
  \centering
  \includegraphics[width=0.78\linewidth]{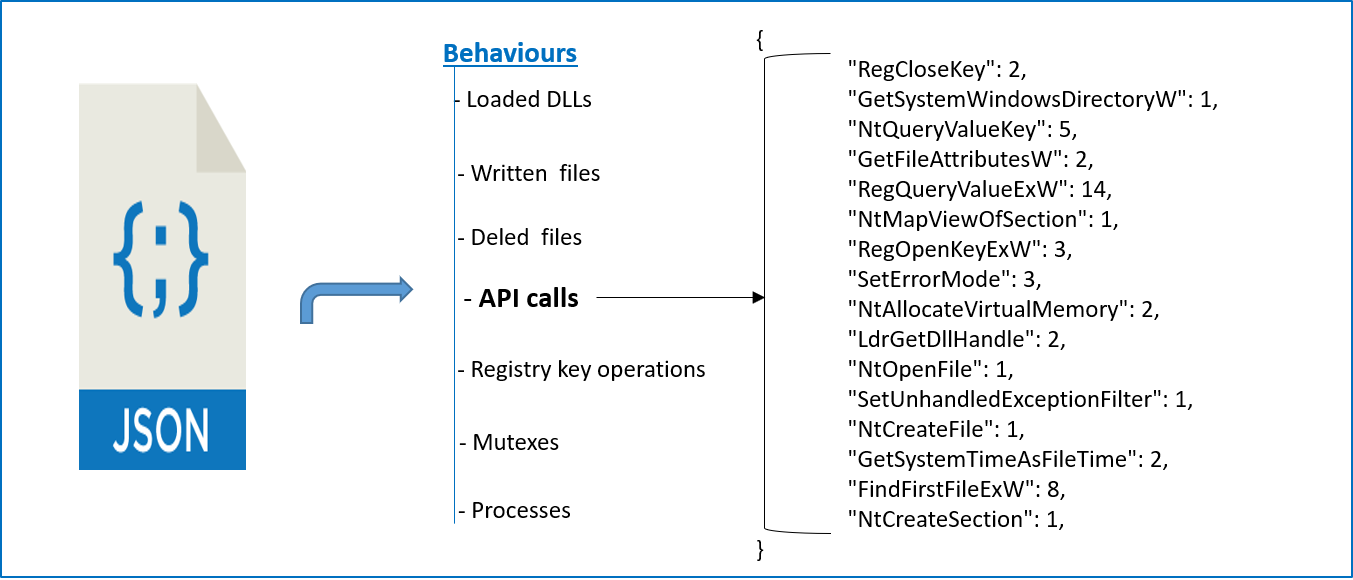}
  \caption{Example of API calls from JSON report of ae03e1079ae2a3b3ac45e1e360eaa973.virus malware captured during dynamic analysis.}
  \label{json-api}
\end{figure*}

The architecture of the analysis environment is mainly based on nested virtualization technology which allows deploying nested virtual machines in the same host. It is also important to note the hardware virtualization on each VirtualBox was set to “enable nested paging” while the paravirtualization interface was set to default. After installing all required software in the Windows 7 VMs, a clean snapshot of each virtual machine was taken and saved. This snapshot is used to restore the virtual machine to its clean state after the analysis of each executable program file. Cuckoo sandbox was configured to generate a behavioural report of each EXE file in JSON format, and each report was further processed to extract relevant features of API calls. A typical structure of JSON report from our dynamic analysis is presented in Figure \ref{json-api} while details on a complete analysis workflow are given in Figure \ref{submitmal} which describes how each file is monitored by the Cuckoo sandbox at runtime.
.

Accordingly, the behavioural reports generated during our dynamic malware analysis reveal that some sophisticated malware can use different API calls that potentially lead to malicious activities. For instance, Table \ref{malicious-api} presents some of the API calls used by ae03e1079ae2a3b3ac45e1e360eaa973.virus, which is a ransomware variant. This ransomware ends up locking the comprised device (Windows VM in our analysis) and demands for a ransom to be paid in BitCoins in order to get access back to the system. This variant encrypts the hard drive’s contents and makes them inaccessible. We have also observed that recent malware variants possess multiple behaviours and can perform multiple malicious activities after compromising the target, making detection difficult.

\begin{table*}[!h]
\caption{Example of potentially malicious API calls observed in ae03e1079ae2a3b3ac45e1e360eaa973.virus while running in a Windows VM during our analysis.}
\label{malicious-api}
\centering
\label{classrresults}
\scalebox{0.99}{
\begin{tabular}{|p{35mm}|p{110mm}|}
\hline
Malicious API call & Description of API Call \\ \hline
WriteConsoleW & Malware uses this API call to establish a command line console. \\ \hline
NtProtectVirtualMemory   &  This API call was used by the malware to allocate read-write memory usually to unpack itself.    \\ \hline
CreateProcessInternalW  &  A process created a hidden Windows to hide the running process from the task manager. This API call allows the malicious program to spawn a new process from itself which is usually observed in packers that use this method to inject actual malicious code in memory and execute them using CreateProcessInternalW.\\ \hline
Process32FirstW  &  The malware used this API to search running processes potentially to identify processes for sandbox evasion, code injection, or memory dumping/live image capturing. \\ \hline
  FindWindowA & The malware used this API to check for the presence of known Windows forensics tools and debuggers that might be running in the background. \\ \hline
\end{tabular}%
}
\end{table*}

\subsection{API Calls Pre-processing Module}
\label{processing-module}
The pre-processing of the behavioural report for each analyzed EXE file is performed at this stage. The pre-processing involves organizing and cleaning the raw features from the behavioural reports to make them suitable for training and testing the classification algorithm. After generating all JSON behavioural reports, they are stored in a local directory in the host machine. A typical JSON report contains various objects storing different data and all behavioural features are located under the “Behaviours” object in each JSON report. However, we are only interested in extracting sequences of API calls of benign and malware executable program files. As JSON stores raw features of API calls, they are not supported by deep learning algorithms. Hence, they must be processed to obtain a numerical representation of each PE file's API call features that is supported by deep learning models.

\begin{figure*}[!ht]
  \centering
  \includegraphics[width=0.87\linewidth]{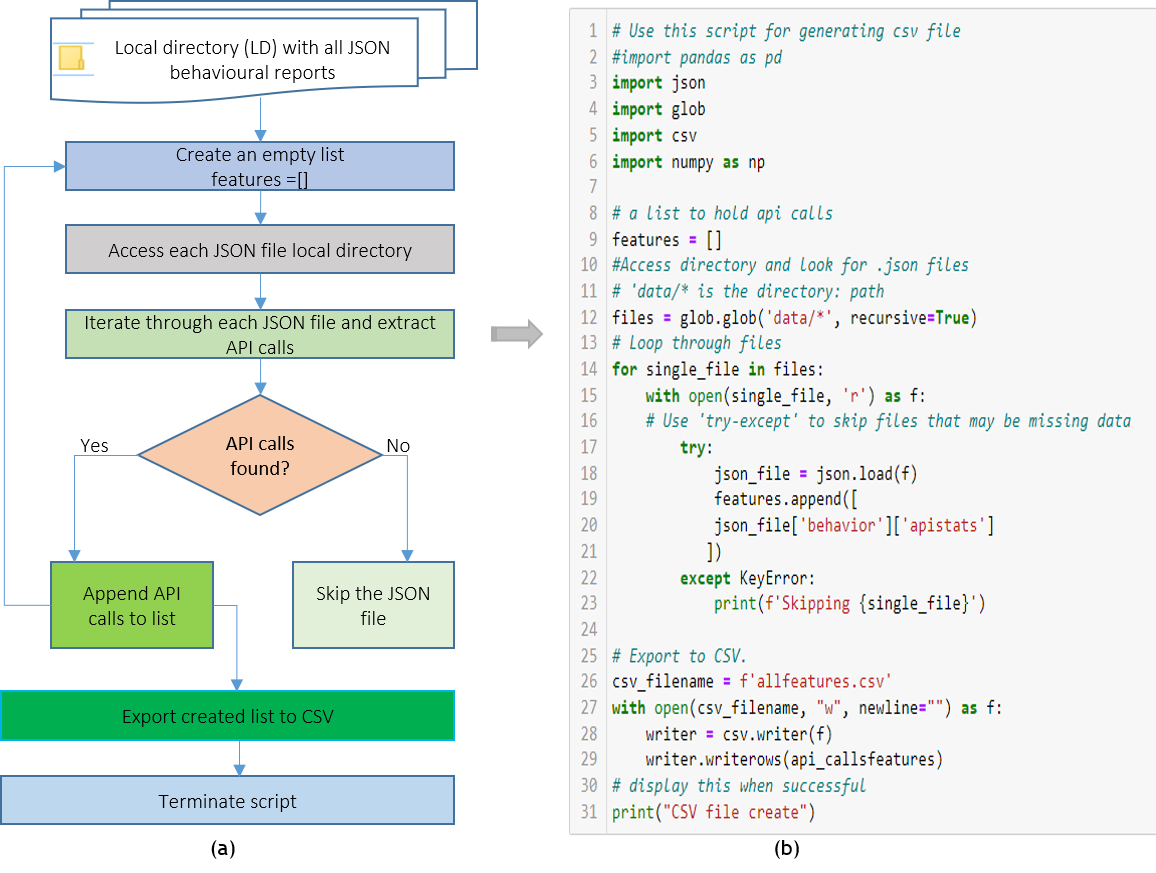}
  \caption{Pre-processing JSON behavioural reports (part 1) with (a) Showing necessary steps and (b) Python code snippet to process JSON file.}
  \label{processingpart1}
\end{figure*}

 The processing involves two main stages. The first stage involves processing all JSON reports to generate a comma separate value (CSV) file. Figure \ref{processingpart1} presents various steps that are performed to accomplish the first processing stage. The second stage deals with processing the CSV file to generate a numerical representation of each feature and construct the embedding matrix which is fed to the proposed CNN-BiGRU hybrid automatic feature extractor. When a CSV file containing sequences of API calls that represent benign or malware program files is fed to the proposed framework, the pre-processing module transforms this CSV file into a pandas data frame which is much easier to process.  A list of labels that are encoded in binary (1,0), with 1 representing each row of malware and 0 representing each row of benign is loaded and then appended to the data frame to obtain a new data frame with labeled features. 

\begin{figure*}[!ht]
  \centering
  \includegraphics[width=0.87\linewidth]{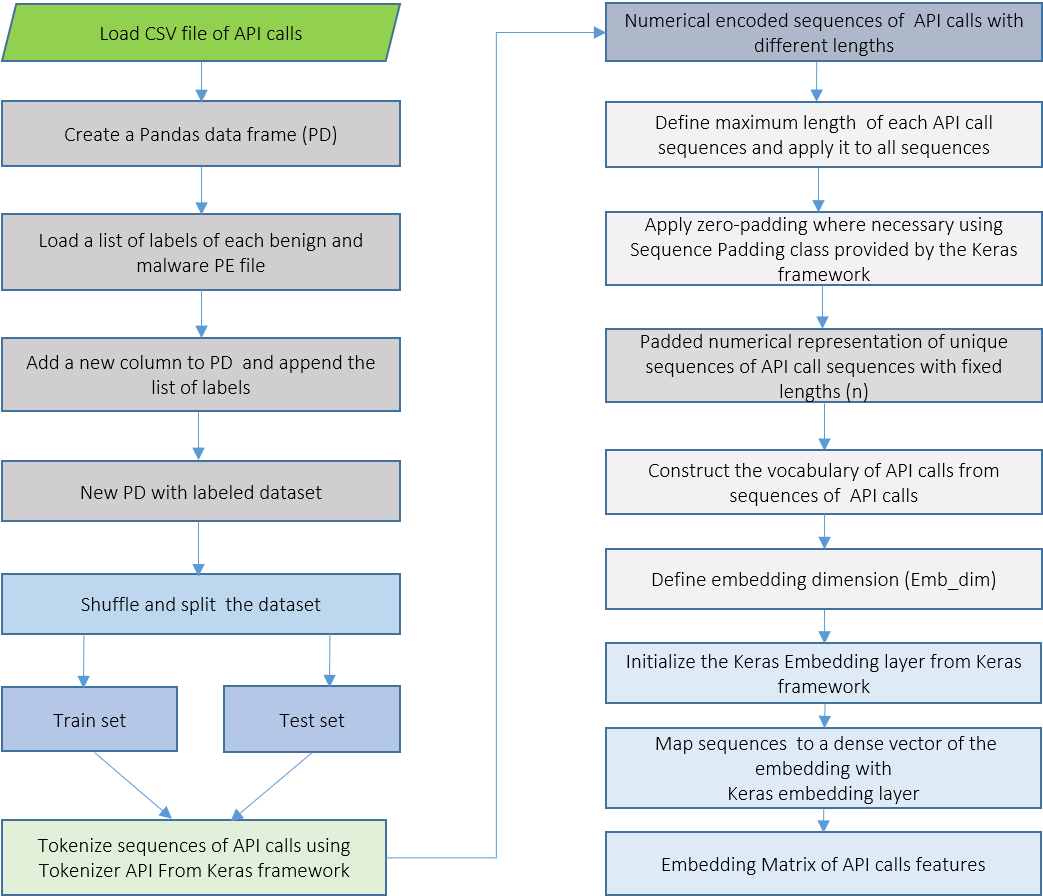}
  \caption{Pre-processing part 2-steps for generating numerical representations of  API calls (encoding each API call).}
  \label{processingpart2}
\end{figure*}

 In the next step, the data are shuffled and passed to a splitting function which splits the dataset into training and test sets. All sequences of API calls (in both training and test set) are tokenized using the Tokenizer API from the Keras framework to generate a unique integer encoding of each API call. Thereafter, all encoded sequences are given the same length and padded where necessary (zero padding is used). This is very important for CNN as it requires all inputs to be numeric and all sequences to have the same length. The encoded sequences are fed to the embedding layer which builds a dense vector representation of each API call and then all vectors are combined to generate an embedding matrix which is used for CNNs. More specifically, all steps performed in the second pre-processing stage are presented in Figure \ref{processingpart2} and further details are provided in subSection \ref{classification}. Additionally, details on the API call embedding are presented in Section \Ref{embedding layer}.

\subsection{Embedding Module}
\label{embedding layer}
This section first introduces word embedding and then provides details on the Keras embedding layer which is used to construct embedding vectors of API calls in this work.
\subsubsection{Word Embedding}
In NLP, word embedding is a group of techniques used to map sequences of words/sentences in a given vocabulary into numerical dense vectors supported by most of the deep learning models\cite{birunda2021review} \cite{rong2016word2vec} and \cite{mikolov2013efficient}. Word embedding techniques reduce the dimensionality of text and can produce features that are more relevant or informative compared to the original raw dataset from which they are generated \cite{chandak2021}. Count-based models and predictive models (often referred to as neural probabilistic language models) are two main categories of word embedding \cite{mandelbaum2016word}. Counted-based models such as Count vectorization and term frequency-inverse document frequency (TF-IDF) represent word features into vectors with extremely high dimensions (which creates sparse vectors and time complexity) while in predictive models, word features are represented in small learned and dense embedding vectors. Word2Vec is one of the best predictive word embedding models for word representation \cite{mikolov2013efficient} \cite{chandak2021}. Word embeddings have gained popularity due to their frequent use as the first processing layer that processes word encoded features in deep learning models \cite{RecentTr81:online}. This success was mainly achieved based on continuous bag-of-words (CBOW) and skip-gram models, which are two Word2Vec embedding models that produce efficient high-quality distributed word feature vector representations where words with similar meanings are grouped based on their semantic/contextual relationships \cite{mikolov2013efficient}. Categorizing word vectors based on their meaning produce an initial correlation of word features for deep learning models.

Word embedding models are often used when dealing with NLP classification problems. For instance, some recent works have applied Word2Vec in modeling text and documents, classifying sentiments and learning semantic information in DNS \cite{pittaras_2020} \cite{MUHAMMAD2021728} \cite{NAWANGSARI2019360} \cite{STEIN2019216} \cite{LOPEZ2020103823}  \cite{ENRIQUEZ20161}. Hence, our API calls feature vector representation approach is linked to these studies, however, in this work, we do not use existing pre-trained NLP-based word embedding models cite{mikolov2013efficient} \cite{pennington2014glove} because word similarities in API call sequences are dissimilar with ordinary English. Thereby, we use the direct embedding layer provided by the Keras DL framework to automatically learn and generate API calls embedding vectors. The direct embedding allows the knowledge to be incorporated inside the detection model, i.e., the whole knowledge of MalDeConv is incorporated in one component, which is different from the previous techniques). 

\subsubsection{Keras Embedding Layer}
As the Keras embedding layer can be used to train embeddings while building the classification model \cite{brownlee2017deep} \cite{Embeddin90:online}, we have used it to generate embedding vectors of API Calls. The word embedding layer in Keras provides an efficient way to generate word embeddings with dense representation where similar words have similar encoding (i.e., this layer can be used to represent both words and sequences of words using dense vector representations). Therefore, the embedding layer learns the embedding of API calls during model training. We treat each API call sequence extracted from each benign and malware EXE file as a sentence. The embedding layer requires all input of API calls to be integer encoded, the reason why each API call was represented by a unique integer using the Keras API Tokenizer during the pre-processing steps (see Section \ref{processing-module}).
Therefore, the embedding layer is first initialized with random weights, and thereafter, it learns the embedding of all API calls in the dataset. To achieve this, the Keras embedding layer uses the continuous bag-of-words model to build embedding vectors of each API call and requires three parameters to be specified. These parameters include $Input_dim$ which specifies the vocabulary size in the API calls data. For instance, if the API calls data has integer encoded values between 0-1000, then the vocabulary size would be 1001 API calls. The second parameter is the $output_dim$ which denotes the size of the vector space in which words/tokens of API calls are embedded (i.e., it specifies/defines the size of each word’s output vector from the embedding layer). For example, it can be of dimensions 10, 20, 100, or even larger. In this work, different values of $out_dim$ are tested to obtain the suitable dimension of the output. Hence, this value is empirically chosen after testing different values and in our case, we have set the $output_dim value$ to 100. Finally, the last parameter is the $input_length$, which is the length of the input of each API call sequence. For instance, if a sequence of API calls extracted from a malware file has 60 API calls, its input length would be 60. As there are many sequences with different lengths in the dataset, all sequences are processed to have the same $input_length$, value. The embedding layer concatenates all generated embedding vectors of API calls to form an embedding matrix which is used as input to the hybrid feature extractor.

\subsection{Hybrid Automatic Feature Extraction Module}
The next component of the proposed framework is the CNN-BiGRU hybrid automatic feature extractor, and its architecture is presented in Figure\ref{maldetconv} and \ref{cnn-bigru module}. More formally, our binary classification problem of benign and malware’s API call sequences with MalDetConv can be addressed as follows.

\subsubsection{CNN feature extractor}
\label{cnn module}
The CNN API feature extractor is designed based on the CNN text classification model presented in Kim’s work \cite{kim2014convolutional}. As it can be viewed from Figure \ref{maldetconv}, the architecture of CNN is made up of two main components, namely, the convolutional layer and the pooling layer. Below we discuss how the CNN feature extractor is designed, the function of each component, and how they interact to perform the extraction of high-level features of API calls.  Given a sequence $s$ of API calls extracted from the behavioural report of benign or malware, let’s $X_{i\in \mathbb{R}^{d}}$ denotes a $d$-dimensional API call vector representing the $i-th$ API call in $s$ where  $d$ is the dimension of the embedding vector.
Therefore, a sequence $S$ consisting of $n$ API calls from a single JSON report can be constructed by concatenating individual API calls using the expression in \eqref{embvec} where the symbol $\oplus$ denotes the concatenation operator and $n$ is the length of the sequence.

\begin{equation}
\label{embvec}
X_{i:n} =x_{1}\oplus x_{2}\oplus  x_{3}\oplus x_{4}\oplus.......\oplus x_{n}
\end{equation}

We have padded sequences (where necessary) using zero padding values to generate API calls matrix of $k_n$ dimensions having $k$ number of tokens of API call with embedding vectors of length $n$. Padding allows sequences to have a fixed number of $k$ tokens (the same fixed length is kept for all sequences) which is very important as CNN cannot work with input vectors of different lengths. We have set $k$ to a fixed length. To identify and select high relevant/discriminative features from raw-level features of API calls word embedding vectors, the CNN feature extractor performs a set of transformations to the input sequential vector $X_{i:n}$ through convolution operations, non-linear activation, and pooling operations in different layers. These layers interact as follows. 

The convolutional layer relies on defined filters to perform convolutional operations to the input vectors of API calls. This allows the convolutional layer to extract discriminative/unique features of API call vectors that correspond to every filter and feature map from the embedding matrix. As the CNN convolutional filters extract features from different locations/positions in the embedding vectors (embedding matrix), the extracted features have a lower dimension compared to the original sequences/features. Hence, mapping high dimensional features to lower-dimensional features while keeping highly relevant features (i.e, it reduces the dimension of features). Positions considered by filters while convolving to the input are independent for each API call and semantic associations between API calls that are far apart in the sequences are captured at higher layers. We have applied a filter $W\in\mathbb{R}^{m\times n}$ to generate a high-level feature representation, with $m$ moving/shifting over the embedding matrix based on a stride $t$ to construct a feature map $c_{i{}}$ which is computed using the expression in \eqref{poolingmap}. It is important to mention that the multiplication operator (*) which is in the equation \eqref{convolution} denotes the convolutional operation (achieved by performing element wise multiplication) which represents API call vectors from $X_{i{}}$  to $X_i+m-1$ (which means $m$ rows at a time) from $X$ which is covered by the defined filter $W$ using the stride. To make the operation faster, we have kept stride to a value of 1, however, various strides can be adapted as well. Moreover, in the equation \eqref{convolution}, the bias value is denoted by $b_{i{}}$.
 
\begin{equation}
\label{convolution}
C_{i}=f(W*X_{i:i+m-1+bi})
\end{equation}
 
CNN supports several activation functions such as hyperbolic tangent, Sigmoid, and rectified linear unit (ReLU). In this work, we have used ReLU, which is represented by $f$ in \eqref{convolution}. Once applied to each input $x$, the ReLu activation function introduces non-linearity by caping/turning all negative values to zero. This operation is achieved using the expression in \eqref{activationrelu}. This activation operation speeds up the training of the CNN model, however, it does not produce significant difference in the model's classification accuracy.

\begin{equation}
\label{activationrelu}
f(x)=max(0,1) 
\end{equation}
 

After convolving filters to the entire input embedding matrix, the out is a feature map corresponding to each convolutional operation and is obtained using the expression in \eqref{poolingmap}.
 
 \begin{equation}
\label{poolingmap}
 C(f) =[C_{1}, C_{2}, C_{3},...,C_{n-m+1}]
\end{equation}

The convolutional layer passes its output to the pooling layer which performs further operations to generate a new feature representation by aggregating the received values. This operation is carried out using some well-known statistical techniques such as computing the mean or average, finding the maximum value, and applying the L-norm. One of the advantages of the pooling layer is that, it has the potential to prevent the model’s overfitting, reducing the dimensionality of features and producing sequences of API call features with the same fixed lengths. In this work, we have used max pooling \cite{kim2014convolutional} \cite{collobert2011natural} which performs the pooling operation over each generated feature map and then selects the maximum value associated with a particular filter output feature map. For instance, having a feature map $c_{i{}}$ , the max-pooling operation is performed by the expression in \eqref{maxpooling} and the same operation is applied to each  $c_{i{}}$ feature map. 
\begin{equation}
\label{maxpooling}
\hat{c}_{i} = max(c_{i}) 
\end{equation}

The goal is to capture the most high-level features of API calls (the ones with the maximum/highest value for every feature map). Note that the selected value from each feature map corresponds to a particular API call feature captured by the filter while convolving over the input embedding matrix. All values from the pooling operations are aggregated together to produce a new reduced feature matrix which is passed to the next feature extractor (BiGRU). 


\subsubsection{BiGRU feature extractor}
\label{bigru-extr}
Features generated by CNN have a low-level semantic compared to the original ones. Fortunately, gated recurrent units can be applied to the intermediate feature maps generated by CNN. Hence, we use the BiGRU module to capture more relevant features of API calls, i.e., the final features maps $\hat{c_{i}}$ of API calls generated by the CNN feature extractor are fed to the BiGRU feature which models sequences in both directions, allowing the model to capture high dependencies across the API features maps produced by CNN. The out consists of relevant information/features which are passed to a flatten layer. 

\subsubsection{Flatten layer}
\label{flatten-layer}
The flatten layer is used to convert/flatten the multi-dimensional input tensors  $\hat{t_{i}}$ produced by the BiGRU automatic feature extractor into a single dimension (a one-dimensional array/vector) which is used as input to the next layer. That is, the output of the BiGRU is flattened to create a single feature vector that is fed to the fully connected neural network module for malware classification.

\subsubsection{Classification Module}
\label{fcnns}
 In our case, the classification module consists of a fully connected neural network (FCNN)/an artificial neural network’s component with hidden layers and a ReLU activation function, and finally the output layer with a sigmoid activation function. The hidden layer neurons/units receive the input features $\hat{t_{i}}$ from the flattening layer and then compute their activations $l_{i{}}$ using the expression in \eqref{fcnnsforward} with $W$ being the matrix of weights between the connections of the input neurons and hidden layer neurons while $b_{i{}}$ represents the biases. We have used the dropout regularization technique to prevent the FCNNs from overfitting the training data, i.e., a dropout rate of 0.2 was used after each hidden layer, which means that at each training iteration, 20\% of the connection weights are randomly selected and set to zero. Dropout works by randomly dropping out/disabling neurons and their associated connections to the next layer, preventing the network’s neurons from highly relying on some neurons and forcing each neuron to learn and to better generalize on the training data \cite{srivastava2014dropout}.

  
\begin{equation}
\label{fcnnsforward}
l_{i}=ReLU(\sum_{i}^{}W_{i}*\hat{t_{i}}+b_{i{}})
\end{equation}

In addition, we have used the binary-cross entropy \cite{jadon2020survey} to computer the classification error/loss and the learning weights are optimized by Adaptive Moment Estimation (Adam) optimizer \cite{kingma2014adam} \cite{yaqub2020state}. Cross-entropy is a measure of the difference between two probability distributions for a given set of events or random variables and it has been widely used in neural networks for classification tasks. On the other hand, Adam works by searching for the best weights $W$ and bias $b$ parameters which contribute to minimizing the computed gradient from the error function (binary-cross entropy in this case). Note that the network learning weights $W$ are updated through backpropagation operations. For the sigmoid function, we have used the binary logistic regression (see equation \eqref{sig}.

\begin{equation}
\label{sig}
Sigmoid (x) =\frac{1}{1+e^{-x}}
\end{equation}
\begin{equation}
\label{classoutcome}
Y=Sigmoid(W_{i}\times l_{i}+b)
\end{equation}

\begin{figure*}[!ht]
  \centering
  \includegraphics[width=0.93\linewidth]{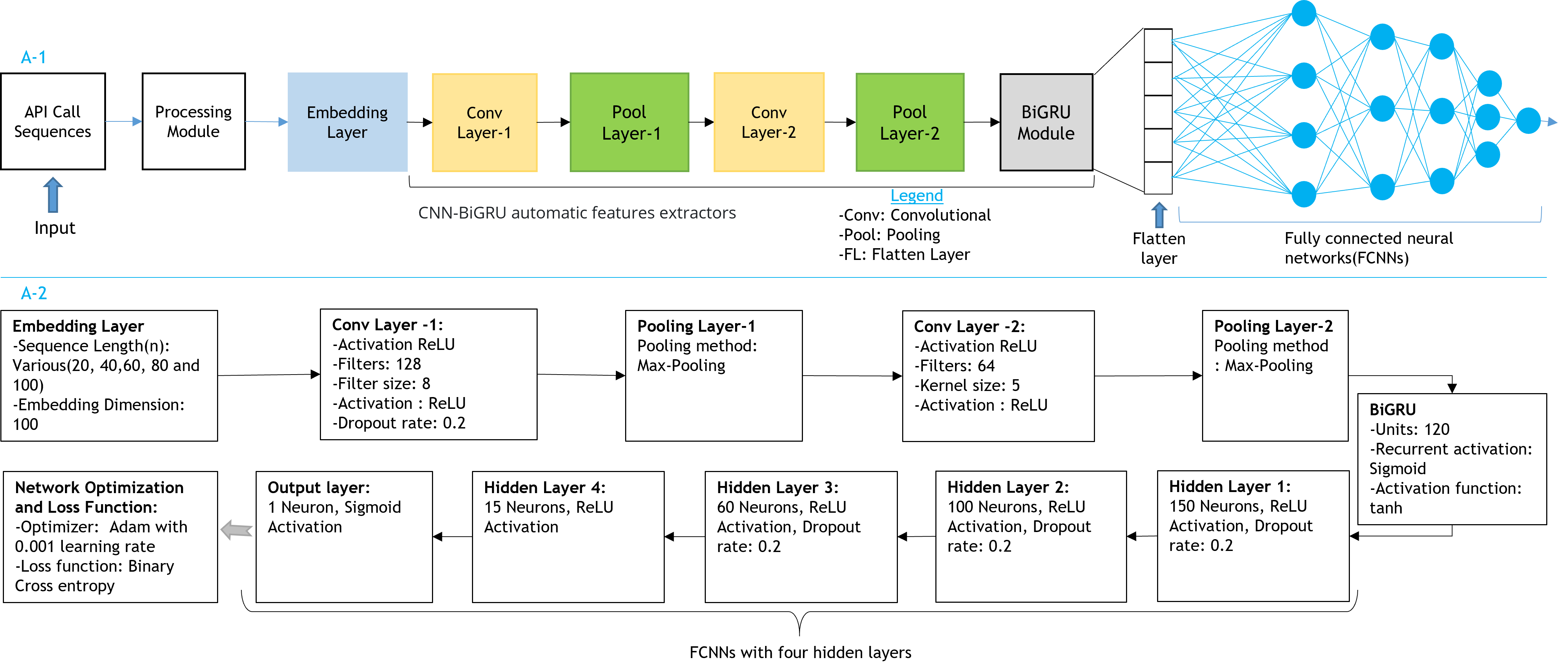}
  \caption{The architecture of the Proposed (A-1) CNN-BiGRU automatic feature extractors (A-2)  Network parameters.}
  \label{cnn-bigru module}
\end{figure*}

After computing the activation, the classification outcome (the output) is computed by the above expression in \eqref{classoutcome}. In addition, a simplified architecture of the proposed hybrid automatic feature extractor and classification module is presented in Figure \ref{cnn-bigru module}. Our CNN network has 2 convolutional layers and 2 pooling layers which are connected to a BiGRU layer. The best parameters for the filter size were determined using the grid search approach. The next component to the BiGRU layer is the FCNN (with hidden layers and output layer) which classifies each EXE file as benign or malicious based on extracted features of API calls (see Figure\ref{cnn-bigru module}(A-1). All network parameters and configurations at each layer are presented in Figure \ref{cnn-bigru module} (A-2).

 \begin{figure*}[!ht]
  \centering
  \includegraphics[width=0.93\linewidth]{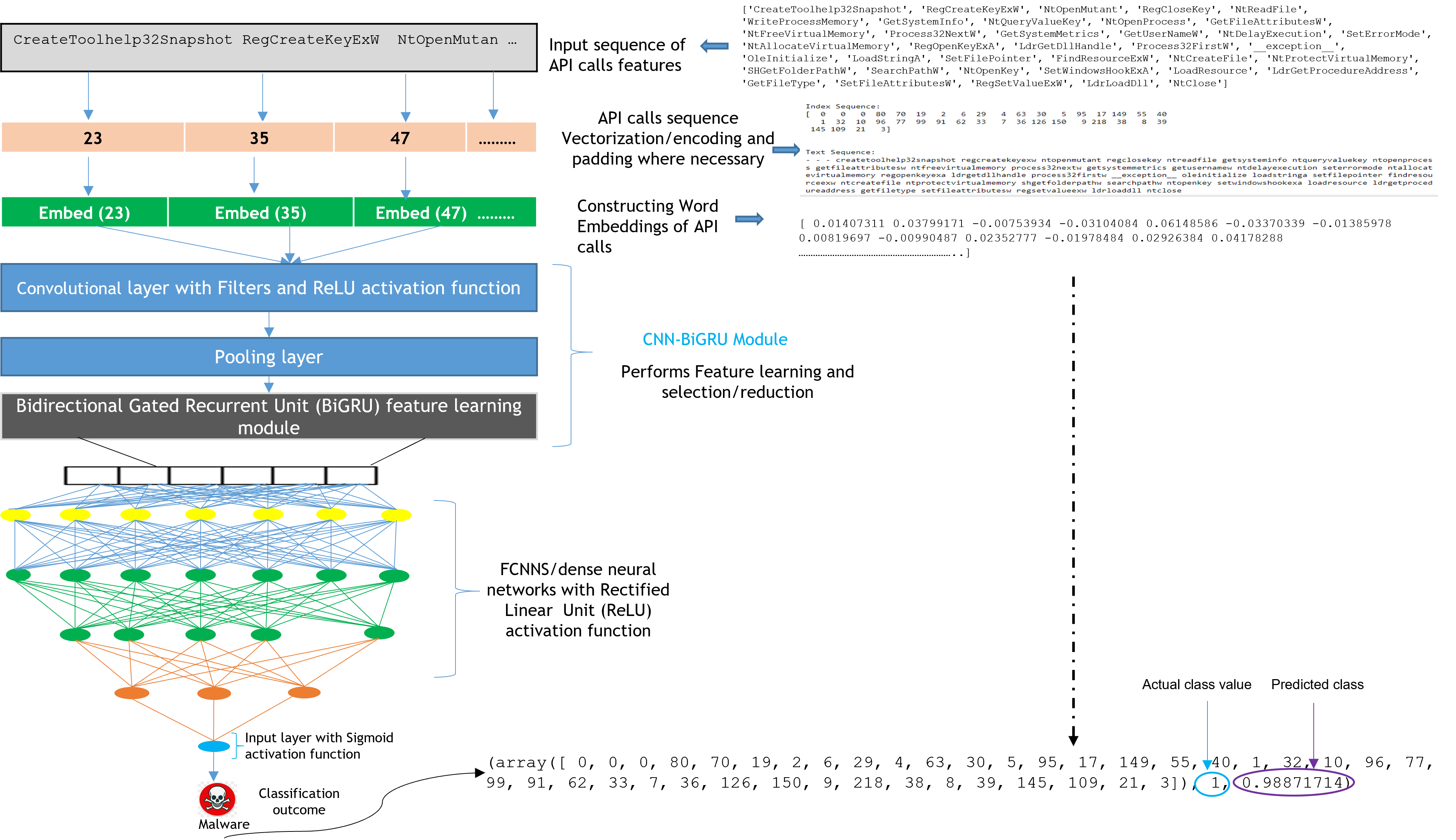}
  \caption{A simplified workflow depicting how MalDetConv works to classify benign and malware executable files' activities as normal or malicious.}
  \label{maldetclass}
\end{figure*}

\subsection{Malware detection with MalDetConv}
\label{classification}
Given a JSON report containing features of API call sequences representing benign or malware  executable programs, the proposed framework performs classification based on all steps described in the previous Sections. First, each JSON file is processed to extract raw sequences of API call features which are followed by encoding each sequence of API calls to obtain an encoded integer representation for each API call. Thereafter, all encoded features are passed to the embedding layer which builds embedding vectors and then concatenates them to generate API calls embedding matrix which is passed to the CNN-BiGRU hybrid automatic feature extractor.  The CNN module has filters that convolve to the embedding matrix to select high-level features through convolution operations. The selected features are aggregated together to generate a feature map which is passed to the pooling layer to generate high-level and compact feature representation through max pooling operations. The output from the last pooling layer is passed to a BiGRU layer to capture dependencies between features of API call and constructs more relevant features which are flattened by the flatten layer and then passed to a fully connected neural network module which performs classification. A complete workflow is depicted in Figure \ref{maldetclass}.

\begin{table*}[]
\centering
\caption{Different datasets of API calls for our experimental analysis.}
\label{dataset-all}
\scalebox{0.99}{
\begin{tabular}{|p{40mm}|p{35mm}|p{35mm}|p{25mm}|}
\hline
Dataset of API Calls                                                           & No. Malware Samples & No. Benign Samples & Released in \\ \hline
MalBehavD-V1\cite{mpascoMa95:online} & 1285                & 1285               & 2022         \\ \hline
Allan and John \cite{PDFWindo91:online}            & 452                 & 101                & 2019         \\ \hline
Brazilian \cite{ceschin2018need} \cite{GitHubfa61:online} & 16315  & 5638 & 2018 \\ \hline
Ki-D \cite{ki2015novel} and \cite{GitHuble77:online}      & 23,080 & 300  & 2015 \\ \hline
\end{tabular}%
}
\end{table*}

\section{Experimental Results and Discussion}
\label{result-discus}
Various experimental results are presented and discussed in this Section. The results presented in this section are based on a binary classification problem of benign and malware executable program files in the Windows systems.

\subsection{Experimental Setup and Tools}
\label{lab-env}
The proposed framework was implemented and tested in a computer running Windows 10 Enterprise edition (64bit) with Intel(R) Core (TM) i7-10700 CPU @ 2.90GHz, 16.0 GB RAM, NVIDIA Quadro P620, and 500 GB for the hard disk drive. The framework was implemented in Python programming language version 3.9.1 using TensorFlow 2.3.0 and Keras 2.7.0 frameworks and other libraries such as Scikit-learn, NumPy, Pandas, Matplotlib, Seaborn, LIME, and Natural Language Toolkit (NLTK) have been also used.  All these libraries are freely available for public use and can be accessed from PiPy \cite{The95:online}, the Python package management website. The proposed framework was trained and tested using sequences of API call features extracted from Windows EXE files.

\subsection{Training and Testing Dataset}
\label{data-used}
Our generated dataset (MalbehavD-v1) was employed to evaluate the performance of the MalDetConv framework. As we wanted the proposed framework to learn and be tested on a variety of different datasets, we have also collected other existing datasets of malicious and normal API calls for our experimental analysis. These datasets include the ones in \cite{ki2015novel} \cite{GitHuble77:online}, API calls datasets presented in \cite{PDFWindo91:online} and \cite{GitHubfa61:online}. Using all these datasets allows us to assess the performance of the proposed framework while detecting malware attacks.Note that these datasets are freely available for public use. We have used 70\% of each data for training while the remaining portion (30\%) was used for testing and details on each dataset are presented in Table \ref{dataset-all}.

\subsection{Performance Evaluation of MalDetConv}
\label{eval-metrics}
Different metrics such as precision (P), recall (R), F1-Score, and accuracy were measured to evaluate the performance of the proposed framework. The computations of these metrics are presented in equations \eqref{prec}, \eqref{rec}, \eqref{accur}, and \eqref{f1}, with TP, TN, FP, and FN indicating True Positives, True Negatives, False Positives, and False Negatives, respectively. Additionally, we have also measured the execution time taken while training and testing MalDetConv. 

\begin{equation}
\label{prec}
Precision=\frac{\mathrm{TP} }{\mathrm{TP+FP}}
\end{equation}
\begin{equation}
\label{rec}
Recall =\frac{\mathrm{TP} }{\mathrm{TP+FN}} 
\end{equation}
\begin{equation}
\label{accur}
Accuracy=\frac{\mathrm{TP+TN}}{\mathrm{TP+TN+FP+FN}}
\end{equation}
\begin{equation}
\label{f1}
F_1-Score =2 \times \frac{\mathrm{Precision \times Recall}}{\mathrm{Precision+Recall}}
\end{equation}

 
  
 
  
\begin{table*}[!ht]
\centering
\caption{Detection accuracy achieved by MalDetConv onMalbehavD-v1 dataset with different length of API call sequence(n).}
\label{maldetconv-on-malbehavD-v1}
\scalebox{1.06}{ 
\begin{tabular} {|p{35mm} | p{50mm} | p{50mm}|}
\hline
n & Training  Accuracy ( \%) & Testing Accuracy ( \%) \\ \hline
20  & 99.17\%  & 93.77\% \\ \hline
40  & 99.44\%  & 94.03\% \\ \hline
60  & 99.72\%  & 95.19\% \\ \hline
80  & 99.56 \% & 95.45\% \\ \hline
100 & 99.72\%  & 96.10\% \\ \hline
\end{tabular}%
}
\end{table*}

\begin{table*}[]
\centering
\caption{MalDetConv peformance on MalbehavD-v1 dataset with different length of API call sequences(n).}
\label{maldetconv-other-metrics}
\scalebox{1.06}{ 
\begin{tabular} {|p{20mm} | p{27mm} | p{27mm} | p{27mm} | p{27mm}|}
\hline
n & Predicted Class & Precision & Recall & F1-Score \\ \hline
\multirow{2}{*}{20}             & Benign          & 0.9128    & 0.9692 & 0.9401   \\ \cline{2-5} 
                                & Malware         & 0.9664    & 0.9055 & 0.9350   \\ \hline
\multirow{2}{*}{40}             & Benign          & 0.9386    & 0.9434 & 0.9410   \\ \cline{2-5} 
                                & Malware         & 0.9420    & 0.9370 & 0.9395   \\ \hline
\multirow{2}{*}{60}             & Benign          & 0.9467    & 0.9589 & 0.9527   \\ \cline{2-5} 
                                & Malware         & 0.9574    & 0.9449 & 0.9511   \\ \hline
\multirow{2}{*}{80}             & Benign          & 0.9515    & 0.9589 & 0.9552   \\ \cline{2-5} 
                                & Malware         & 0.9577    & 0.9501 & 0.9539   \\ \hline
\multirow{2}{*}{100}            & Benign          & 0.9521    & 0.9717 & 0.9618   \\ \cline{2-5} 
                                & Malware         & 0.9705    & 0.9554 & 0.9602   \\ \hline
\end{tabular}%
}
\end{table*}
  
\subsubsection{Classification Results}
\label{class-results}

The MalDetConv framework was trained and tested on different datasets to observe how effective it is while detecting/classifying unknown malicious activities. Hence, various evaluations were carried out, and the classification results are presented in this Section. As mentioned in the previous sub-Section \ref{data-used}, we evaluate MalDetConv on API call sequences extracted from Windows EXE files through the dynamic analysis approach. We have also compared the performance of MalDetConv against other state-of-the-art techniques presented in the previous works, which reveals its effectiveness and ability to handle both existing and newly emerging malware attacks over its counterparts. Accordingly, Table \ref{maldetconv-on-malbehavD-v1} shows the training and testing classification accuracy achieved by MalDetConv on our dataset (MalbehavD-v1). Different lengths (n) of API call sequences were considered to evaluate their effects on the performance. From the results, we could see that MalDetConv successfully classified malicious activities and benign activities with the accuracy of 96.10\% where $n=100$. The lowest accuracy (93.77\%) was obtained using the lengths of 20. Interestingly, the testing accuracy varies in accordance with the value of $n$, revealing the effect of the API call sequence’s length on the performance, i.e., as $n$ increases, the accuracy also increases. This is shown by the accuracy improvement of 2.33\% $(96.10-93.77)$ obtained by increasing the value of $n$ from 20 to 100.


Table \ref{maldetconv-other-metrics} presents the precision, recall, and F1-score obtained while testing MalDetConv, which demonstrates better performance while detecting both malware and benign on unseen data. For instance, the precision of 0.9128 and 0.9664 was obtained using the length of the API call sequence of 20 for benign and malware detection, respectively. Similarly, using the same length (n=20) the MalDetConv achieves the recall of 0.9055 and F1-Score of 0.9350 for malware detection. The precision values obtained using different values of $n$ (20, 40, 60, 80, and 100) demonstrate a better performance of the MalDetConv framework when distinguishing between malware and benign activities (positive and negative classes). That is, our framework has a good measure of separability between both classes (performs well in identifying malware attacks) and can deal with long sequences. The higher the value of precision (with 1 being the highest), the better performance of a given detection model. It is also important to highlight that in many cases the precision increases as $n$ increases.

\begin{figure*}[!ht]
  \centering
  \includegraphics[width=0.88\linewidth]{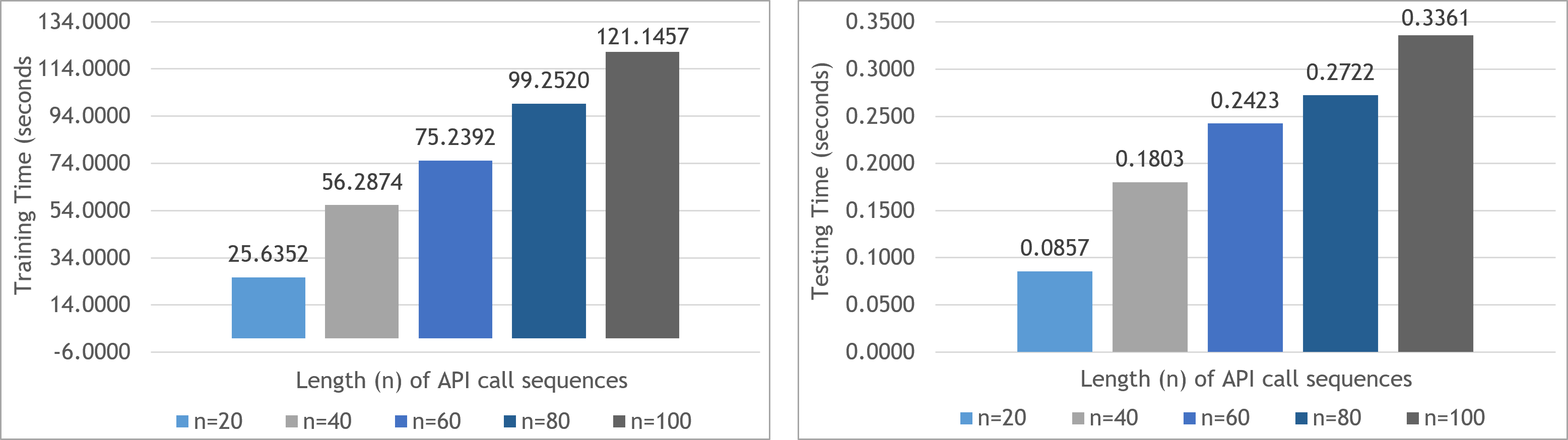}
  \caption{Execution time (a) Training time (b) Testing time taken by MalDetConv framework on our dataset.}
  \label{Execution-maldetconv}
\end{figure*}

Figure \ref{Execution-maldetconv} presents the execution time taken by the MalDetConv framework which shows that increasing $n$ results in a high training time (see Figure \ref{Execution-maldetconv}). However, the testing time is low, and it does not dramatically increase, which again proves the effectiveness of MalDetConv in terms of malware detection time.



\begin{table*}[!h]
\centering
\caption{Comparison of MalDetConv framework against other DL-based techniques using MalbehavD-V1 dataset.}
\label{comparisons-dl}
\scalebox{1.22}{ 
\begin{tabular} {|p{28mm} | p{28mm} | p{28mm} | p{28mm}|}
\hline
Detection Method  & Precision & Recall & F1-Score \\ \hline
MLP               & 0.9459    & 0.9186 & 0.9321   \\ \hline
LSTM              & 0.9491    & 0.9291 & 0.9390   \\ \hline
BiLSTM            & 0.9375    & 0.9449 & 0.9412   \\ \hline
CNN               & 0.9678    & 0.9475 & 0.9576   \\ \hline
GRU               & 0.9521    & 0.9396 & 0.9458   \\ \hline
BiGRU             & 0.9671    & 0.9265 & 0.9464   \\ \hline
MalDetConv        & 0.9705    & 0.9554 & 0.9602   \\ \hline
\end{tabular}%
}
\end{table*}

\begin{figure*}[!ht]
  \centering
  \includegraphics[width=0.86\linewidth]{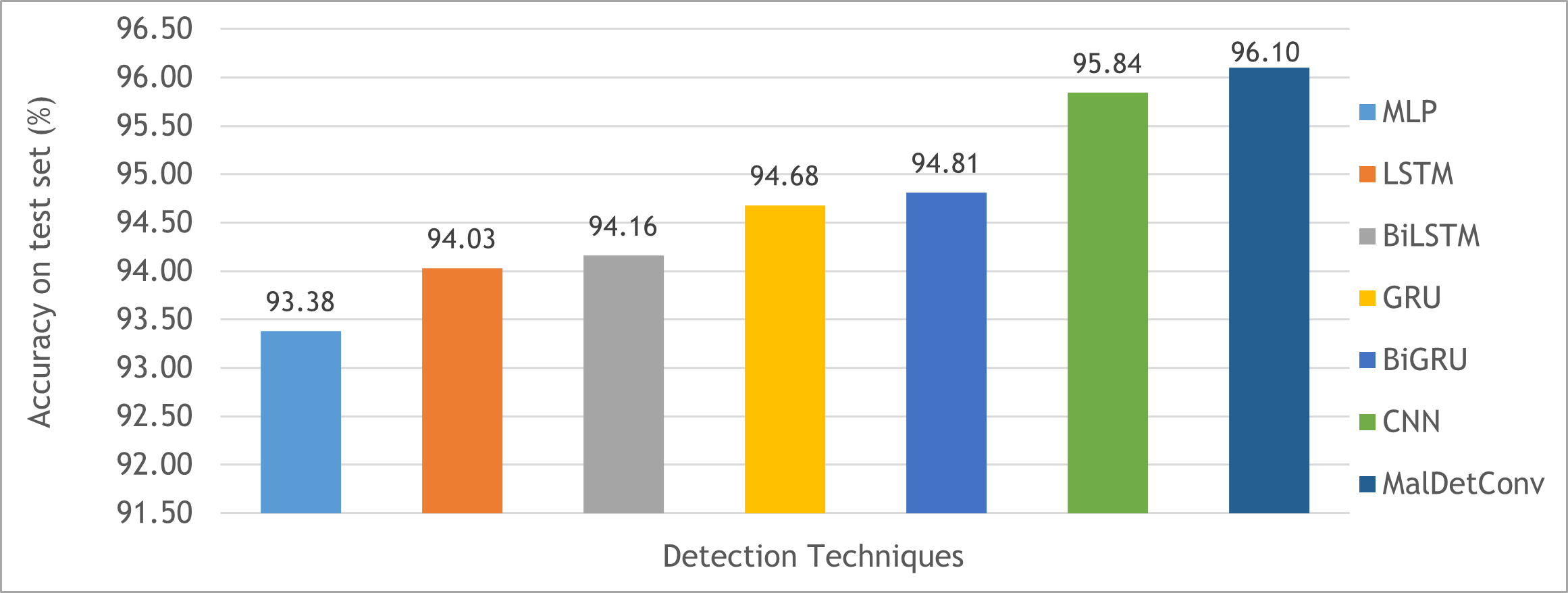}
  \caption{Comparison of MalDetConv framework  against other DL-based techniques in terms of detection accuracy.}
  \label{DL comparison1}
\end{figure*}

We have also implemented other DL models and compared their performance with MalDetConv. The detection results on unseen EXE files (API call sequences) presented in Table \ref{comparisons-dl} and Figure \ref{DL comparison1} indicate better precision, recall and F1-Score, and accuracy of the proposed framework, compared to the existing deep learning models. MalDetConv outperforms multilayer perceptron (MLP), LSTM, BiLSTM, GRU, BiGRU, and CNN using the MalbehavD-v1 dataset. MLP achieves the Accuracy of 93.38\%, LSTM (94.03\%), BiLSTM (94.16\%), and CNN (95.84\%). The highest improvement gap in detection accuracy of 2.72\% (96.1-93.38) is observed between the performance of MLP and MalDetConv.


\begin{table*}[]
\centering
\caption{Performance of MalDetConv framework on other datasets of API call sequences (n=100).}
\label{maldetconv-other-dataset}
\scalebox{1.05}{ 
\begin{tabular} {|p{30mm} | p{26mm} | p{28mm} | p{24mm} | p{24mm} | p{24mm}|} 
\hline
Dataset                          & Predicted Class & Precision & Recall & F1-Score \\ \hline
\multirow{2}{*}{Allan  and John} & Benign          & 1.0000    & 0.7667 & 0.8679   \\ \cline{2-5} 
                                 & Malware         & 0.9504    & 1.0000 & 0.9745   \\ \hline
\multirow{2}{*}{Brazilian}       & Benign          & 0.9621    & 0.9656 & 0.9638   \\ \cline{2-5} 
                                 & Malware         & 0.9884    & 0.9872 & 0.9878   \\ \hline
\multirow{2}{*}{Ki-D}            & Benign          & 1.0000    & 0.9524 & 0.9756   \\ \cline{2-5} 
                                 & Malware         & 0.9993    & 1.0000 & 0.9996   \\ \hline
\end{tabular}%
}
\end{table*}

\begin{figure*}[!ht]
  \centering
  \includegraphics[width=0.88\linewidth]{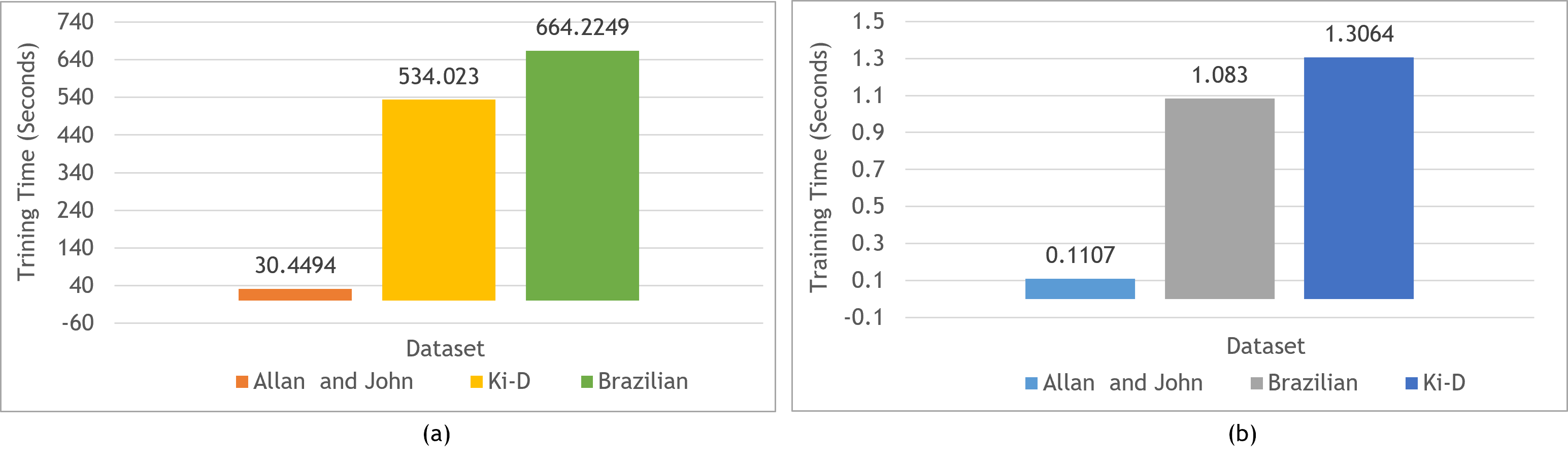}
  \caption{Execution time (a)Training and (b) Testing times taken  by the proposed framework on the previous datasets of API calls.}
  \label{execution-other datasets}
\end{figure*}


\begin{table*}[]
\centering
\caption{Comparing MalDetConv against previous NLP-based techniques/frameworks based on sequences of API calls dataset.}
\label{Comparison-MalDetConv}
\scalebox{0.72}{ 
\begin{tabular} {|p{32mm} | p{37mm} | p{60mm} | p{47mm} | p{28mm}|}
\hline
Dataset &
  Techniques/Framework &
  Feature Vectorization Approach &
  ML/DL  Algorithm &
  Detection   Accuracy \\ \hline
\multirow{2}{*}{Allan and John} & MalDy \cite{karbab2019maldy}        & TF-IDF                 & XGBoost                      & 95.18\%          \\ \cline{2-5} 
                                & MalDetConv    & Word2Vec-CBOW          & CNN-BiGRU                    & \textbf{95.73\%} \\ \hline
\multirow{2}{*}{MalBehavD-V1}   & MalDy \cite{karbab2019maldy}        & TF-IDF                 & XGBoost         &  95.59\%          \\ \cline{2-5} 
                                & MalDetConv    & Word2Vec-CBOW          & CNN-BiGRU                    & \textbf{96.10\%} \\ \hline
\multirow{4}{*}{Brazilian}      & MalDy   \cite{karbab2019maldy}      & TF-IDF                 & KNN                          & 97.60\%          \\ \cline{2-5} 
 &
  Amer et al. \cite{amer2022malware} &
  TF-IDF+word2vec &
  Swarm intelligence algorithm &
  95.4\% \\ \cline{2-5} 
 &
  Ceschin et al. \cite{8636415} &
  TF-IDF &
  Random Forest &
  98.00\% \\ \cline{2-5} 
                                & MalDetConv    & Word2Vec-CBOW          & CNN-BiGRU                    & \textbf{98.18\%} \\ \hline
\multirow{4}{*}{ki-D} &
  Amer and Zelinka \cite{Amer2020} &
  Word embedding and clustering similarity &
  Markov chain model &
  99.90\% \\ \cline{2-5} 
                                & Ki et al.  \cite{ki2015novel}    & DNA Sequence Alignment & API call matching algorithm  & 99.80\%          \\ \cline{2-5} 
                                & Tran and Sato  \cite{8233569}  & TF-IDF                 & Support Vector Machine (SVM) & 96.18\%          \\ \cline{2-5} 
                                & MalDetConv    & Word2Vec-CBOW          & CNN-BiGRU                    & \textbf{99.93\%} \\ \hline
\end{tabular}%
}
\end{table*}

Table \ref{maldetconv-other-dataset} shows the classification results of MalDetConv on existing datasets of API calls (Allan and John, Brazilian, and Ki-D datasets). Using Allan and John’s dataset, MalDetConv detects unseen malware with a precision of 0.9504 and detects benign files’ activities with a precision of 1.0000. A precision of 0.9884 was obtained by MalDetConv on the Brazilian dataset while with the Ki-D dataset, MalDetConv performed with a precision of 0.9993. The detection results presented in Table obtained by measuring all metrics (precision, recall, and F1-score)  \ref{maldetconv-other-dataset}, demonstrate better performance of MalDetConv on both small and large datasets. Looking at the training time presented in Figure \ref{execution-other datasets}, there is an increase in the training time as the size of the dataset increases. For instance, the execution time of 30.4494 seconds was taken while training MalDetConv on Allan and John’s dataset, and the testing time of 0.1107 seconds was taken while testing the framework on unseen samples. In addition, the training time of 664.2249 seconds was taken to train MalDetConv on the  Brazilian dataset. Nevertheless, a few seconds (1.083) were taken to test the model on the test set. The huge gap in the execution time is mainly due to the size of the dataset. It is worth mentioning that the detection results were generated on the test set (30\% of each dataset) and the length of the API call sequence of 100. Despite the training time which tends to be high for large datasets, it is worth mentioning that the MalDetConv framework does not take a huge amount of time to detect malware attacks, which is a necessity for any robust and efficient anti-malware detection system. 


\begin{figure*}[!ht]
  \centering
  \includegraphics[width=0.88\linewidth]{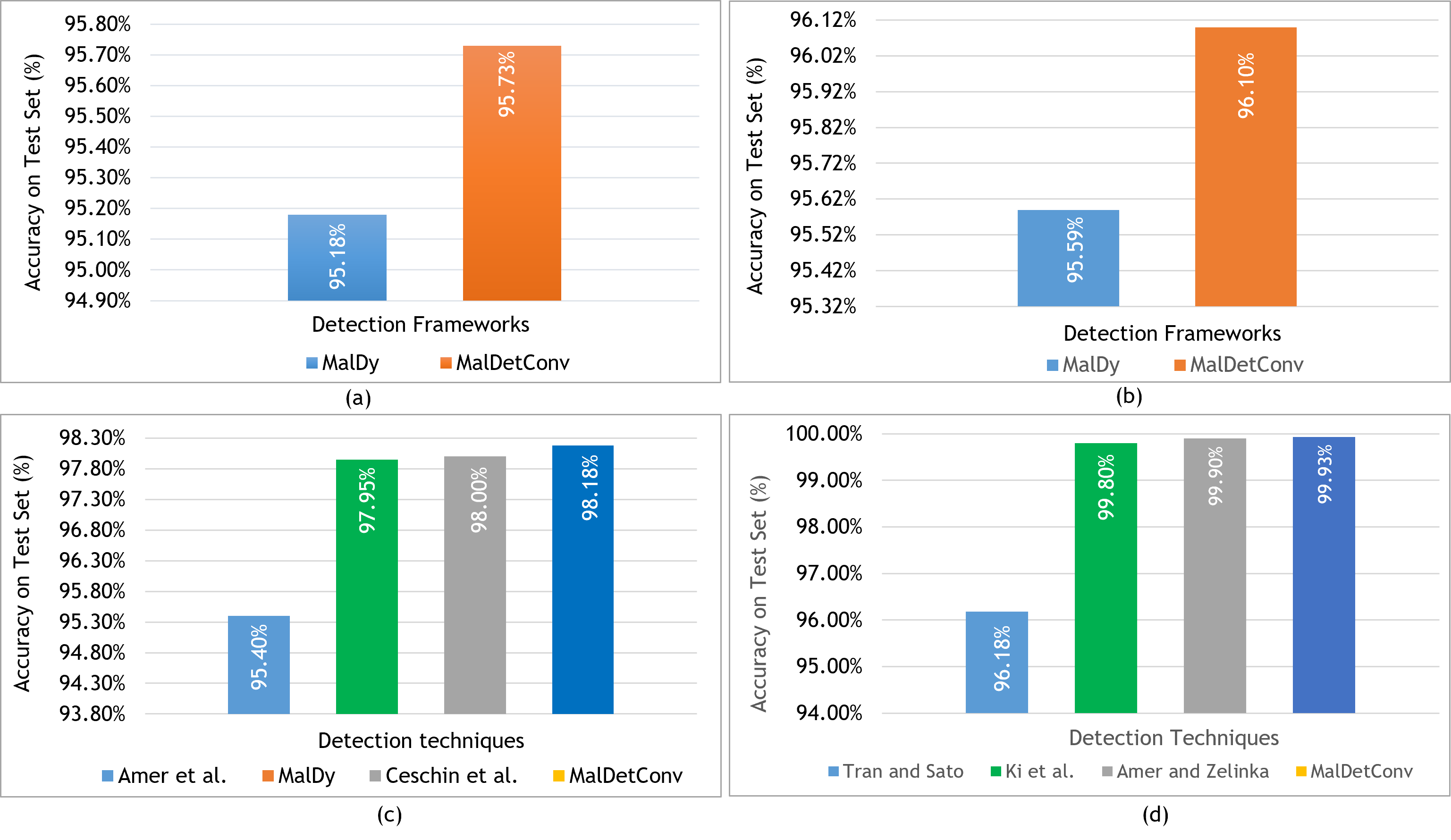}
  \caption{Detection accuracy of MalDetConv against other techniques using (a)  Allan and David (b) MalbehavD-V1 (c) Brazilian and  (d) Ki-datasets \ref{Comparison-MalDetConv}.}
  \label{Maldet2}
\end{figure*}

We have also examined the performance of the MalDetConv framework against other existing NLP-based frameworks/techniques based on API calls extracted from EXE files and comparative results are presented in Table \ref{Comparison-MalDetConv}. First, we compared MalDetConv against Maldy\cite{karbab2019maldy}, an existing framework based on NLP and machine learning techniques. Using our dataset, Maldy achieved the detection accuracy of 95.59\% with XGBoost while MalDetConv achieved 96.10\%, creating an improvement of 0.51\% (96.10\%-95.59\%) detection accuracy made by MalDetConv. The detection improvement of 0.55\% (95.73\% - 95.18\%) was achieved by MalDetConv over the Maldy framework using Allan and John’s dataset. Additionally, MalDetConv detection also outperforms the Maldy on Ki-D and Brazilian datasets. Compared with the malware detection technique presented by Amer and Zelinka \cite{Amer2020}, MalDetConv obtained an improvement of 0.3\% (99.93\% - 99.90\%) on Ki-D dataset.  Moreover, there is an improvement of 0.13\% (99.93\% - 99.80\%) in the detection accuracy achieved over the technique implemented by Ki et al. \cite{ki2015novel} using the Ki-D dataset, 0.58\% (98.18\%-97.60\%) over Ceschin et al.’s \cite{8636415} detection technique achieved using the Brazilian dataset and an improvement of 3.75\% (99.93\%-96.18\%) over the detection technique presented by Tran and Sato \cite{8233569}. MalDetConv has also outperformed the performance of a malware detection technique presented by Amer et al. \cite{amer2022malware}, with an improvement of 2.76\% (98.18\%-95.40\%) in the accuracy achieved using the Brazilian dataset. Figure \ref{Maldet2} also depicts the comparison between MalDetConv against the above techniques. The overall performance achieved in various experiments shows that the MalDetConv framework can potentially identify and detect malware attacks with a high precision and detection accuracy, giving our framework the ability to deal with different malware attacks on Windows platforms.

\begin{figure*}[!ht]
  \centering
  \includegraphics[width=0.85\linewidth]{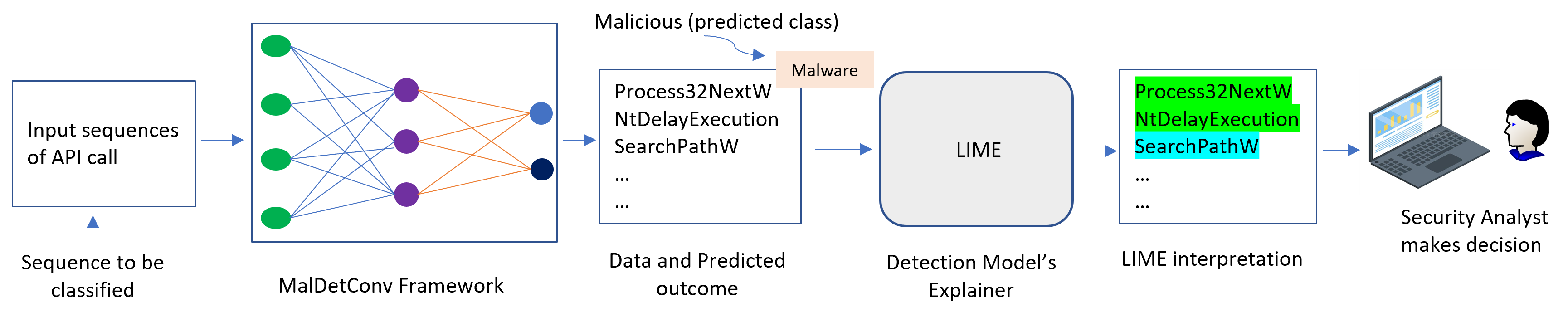}
  \caption{Explaining the predicted outcome. MalDetConv predicts that a sequence of API calls is malicious, and LIME highlights the API calls in the sequence that led/contributed to the prediction. These can help security analysts to make decisions and trust MalDetConv’s prediction.}
  \label{how-lime works}
\end{figure*}

\begin{figure*}[!ht]
  \centering
  \includegraphics[width=0.76\linewidth]{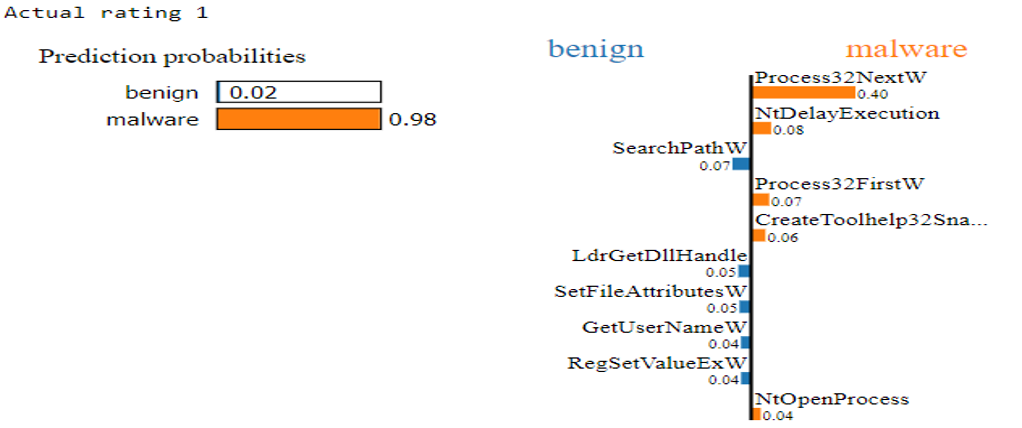}
  \caption{An example of an explanation of the classification outcome generated by LIME when classifying a malware file with MalDetConv.}
  \label{lime output}
\end{figure*}

\subsubsection{Understanding MalDetConv Prediction With LIME}
\label{lime-explanation}
It is often complicated to understand the prediction/classification outcome of deep learning models given the many parameters they use when making a prediction. Therefore, in contrast to the previous malware detection techniques based on API calls, we have integrated LIME \cite{ribeiro2016should} into our proposed behaviour-based malware detection framework which helps to understand predictions. The local interpretable model-agnostic explanations (LIME) is an automated framework/library with the potential to explain or interpret the prediction of deep learning models. More importantly, in the case of text-based classification, LIME interprets the predicted results and then reveals the importance of the most highly influential words (tokens) which contributed to the predicted results.This works well for our proposed framework as we are dealing with sequences of API calls that represent malware and benign files. It is important to mention that the LIME framework was chosen because it is open source, has a high number of citations and the framework has been highly rated on GitHub.

\begin{figure*}[!ht]
  \centering
  \includegraphics[width=0.75\linewidth]{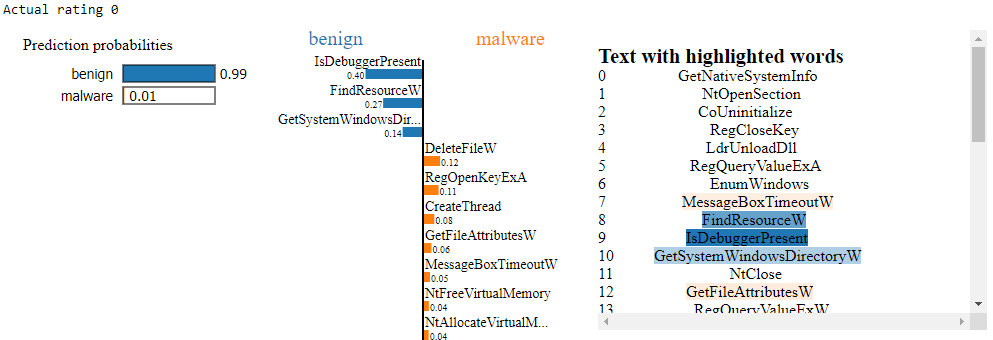}
  \caption{Explanation of the classification outcome generated by Lime when classifying a benign file with the MalDetConv framework.}
  \label{lime output1}
\end{figure*}


Figure \Ref{how-lime works} shows how LIME works to provide an explanation/interpretation of a given prediction of API call sequence. LIME explains the framework's predictions at the data sample level, allowing security analysts/end-users to interpret the framework’s predictions and make decisions based on them. LIME works by perturbing the input of the data samples to understand how the prediction changes i.e., LIME considers a deep learning model as a black box and discovers the relationships between input and output which are represented by the model \cite {ribeiro2016should} \cite{Understa80:online}. The output produced by LIME is a list of explanations showing the contributions of each feature to the final classification/prediction of a given data sample. This produces local interpretability and allows security practitioners to discover which API call feature changes (in our case) will have the most impact on the predicted output. LIME computes the weight probabilities of each API call in the sequence and highlights individual API calls that led to the final prediction of a particular sequence of API calls representing malware or benign EXE file. 


For instance, in Figure \ref{lime output}, the Process32NextW, NtDelayExecution, Process32FirstW, CreateToolhelp32Snapshot, and NtOpenProcess are portrayed as the most API calls contributing to the final classification of the sequence as “malicious” while SearchPathW, LdrGetDllHandle, SetFileAttributesW, and GetUserNameW API calls are against the final prediction. Another example showing Lime output is presented in Figure \Ref{lime output1} where API calls features that led to the correct classification of a benign file are assigned weight probabilities which are summed up to give the total weight of 0.99. API calls such as IsDebuggerPresent, FindResourceW, and GetSystemWindowsDir are among the most influential API calls that contribute to the classification of the sequence/file into its respective class (benign in this case).

The screenshots of lime explanations presented in Figures \ref{lime output} and \ref{lime output1} are generated in HTML as it generates clear visualizations than other visualization tools such as matplotlib.  In addition, the weights are interpreted by applying them to the prediction probability. For instance, if API calls IsDebuggerPresent and FindResourceW are removed from the sequences, we expect MalDetConv to classifier the benign sequence with the probability of 0.99-0.38-0.26= 0.37. Ideally, the interpretation/explanation produced by Lime is a local approximation of the MalDetConv framework’s behaviours, allowing to reveal what happened inside the black box.  Note that in Figure \ref{lime output1}, the tokens under “Text with highlighted words”, represent the original sequences of API while the number before each API call (e.g., 0 for RegCreateKeyExW) corresponds to its index in the sequence.  The highlighted tokens show those API calls which contributed to the classification of the sequence. Thus, having this information, a security analyst can decide whether the model’s prediction should be trusted or not.


\section{Limitations and Future Work}
\label{limitations}
We have covered Windows 7 OS EXE files (malware and benign) in this research work. In our future work, we intend to evaluate the performance of the proposed framework on the latest versions of Windows OS such as Windows 10 or Windows 11 and other Windows file formats such as PowerShell scripts. Our dataset of API calls (MalBehavD-V1) will be extended to include more features from newly released malware variants. We also plan to evaluate the proposed framework on API call features extracted in the Android applications where API calls features will be extracted from APK files through dynamic analysis. Although, currently we can explain the results using LIME, but, in some cases LIME can be unstable as it depends on the random sampling of new features/perturbed features \cite{molnar2020interpretable}. LIME ignores correlations between features as data points are sampled from a Gaussian distribution\cite{molnar2020interpretable}.  Therefore, we plan to explore and compare explanation insights produced by other frameworks such as Anchor\cite{anchors:aaai18} and ELI5 \cite{GitHubTe20:online} which also interpret deep learning models.

\section{Conclusion}
\label{conclu} 
This work has presented a dynamic malware detection framework, MalDetConv. In addition, a new dataset of API call sequences, namely, MalBehavD-V1, are also presented, which is extracted from benign and malware executable program files. The analysis was performed in a virtual isolated environment powered by the Cuckoo sandbox. The proposed framework is mainly based on natural language processing and deep learning techniques. The Keras Tokenizer APIs and embedding layer were used to perform feature vectorization/representation of API call sequences which generate embedding vectors of each API and group them based on their semantic relationships. MalDetConv uses a CNN-BiGRU module to perform automatic extraction and selection of high-relevant features that are fed to a fully connected neural network component for the classification of each sequence of API calls. The performance of MalDetConv was evaluated using the newly generated dataset (MalBehavD-V1) and some of the well-known datasets of API calls presented in the previous works. The experimental results prove the potential of MalDetConv over existing techniques based on API calls extracted from EXE files. The experimental results show high detection accuracy and better precision achieved by MalDetConv on both seen and unseen data. This provides the confidence that this framework can successfully identify and classify newly emerging malware attacks based on their behaviours. Moreover, MalDetConv achieved high detection accuracy on both small and large datasets, making it the best framework able to deal with the current malware variants. To address the problem of model interpretability encountered in most of the deep learning-based techniques, the LIME module was integrated into MalDetConv, enabling our framework to produce explainable predictions. The MalDetConv framework can be deployed in high-speed network infrastructures to detect both existing and new malware in a short time, with high accuracy and precision. 

\bibliography{mybibfile}

\begin{thebibliography}{100}
\expandafter\ifx\csname url\endcsname\relax
  \def\url#1{\texttt{#1}}\fi
\expandafter\ifx\csname urlprefix\endcsname\relax\def\urlprefix{URL }\fi
\expandafter\ifx\csname href\endcsname\relax
  \def\href#1#2{#2} \def\path#1{#1}\fi

\bibitem{cisco2020cisco}
Cisco, Cisco annual internet report (2018--2023) white paper (2020).

\bibitem{IoTconn31:online}
Statista, Iot connected devices worldwide 2019-2030 | statista,
  \url{https://www.statista.com/statistics/1183457/iot-connected-devices-worldwide/},
  (Accessed on 08/04/2022).

\bibitem{maniriho2021study}
P.~Maniriho, A.~N. Mahmood, M.~J.~M. Chowdhury, A study on malicious software
  behaviour analysis and detection techniques: Taxonomy, current trends and
  challenges, Future Generation Computer Systems 130 (2022) 1--18.
\newblock \href {https://doi.org/https://doi.org/10.1016/j.future.2021.11.030}
  {\path{doi:https://doi.org/10.1016/j.future.2021.11.030}}.

\bibitem{ANotSoCo81:online}
B.~Jovanovic, A not-so-common cold: Malware statistics in 2022,
  \url{https://dataprot.net/statistics/malware-statistics/}, (Accessed on
  06/14/2022) (2022).

\bibitem{Over100M58:online}
A.~Drapkin, Over 100 million pieces of malware were made for windows users in
  2021, \url{https://tech.co/news/windows-users-malware}, (Accessed on
  06/14/2022).

\bibitem{zhang2019static}
S.-H. Zhang, C.-C. Kuo, C.-S. Yang, Static pe malware type classification using
  machine learning techniques, in: 2019 International Conference on Intelligent
  Computing and its Emerging Applications (ICEA), IEEE, 2019, pp. 81--86.
\newblock \href {https://doi.org/https://doi.org/10.1109/ICEA.2019.8858297}
  {\path{doi:https://doi.org/10.1109/ICEA.2019.8858297}}.

\bibitem{naik2021fuzzy}
N.~Naik, P.~Jenkins, N.~Savage, L.~Yang, T.~Boongoen, N.~Iam-On, Fuzzy-import
  hashing: A static analysis technique for malware detection, Forensic Science
  International: Digital Investigation 37 (2021) 301139.
\newblock \href {https://doi.org/https://doi.org/10.1016/j.fsidi.2021.301139}
  {\path{doi:https://doi.org/10.1016/j.fsidi.2021.301139}}.

\bibitem{Singh2020a}
J.~Singh, J.~Singh, {Detection of malicious software by analyzing the
  behavioral artifacts using machine learning algorithms}, Information and
  Software Technology 121 (2020).
\newblock \href {https://doi.org/https://doi.org/10.1016/j.infsof.2020.106273}
  {\path{doi:https://doi.org/10.1016/j.infsof.2020.106273}}.

\bibitem{sun2019opcode}
Z.~Sun, Z.~Rao, J.~Chen, R.~Xu, D.~He, H.~Yang, J.~Liu, An opcode sequences
  analysis method for unknown malware detection, in: Proceedings of the 2019
  2nd international conference on geoinformatics and data analysis, 2019, pp.
  15--19.
\newblock \href {https://doi.org/https://doi.org/10.1145/3318236.3318255}
  {\path{doi:https://doi.org/10.1145/3318236.3318255}}.

\bibitem{huda2016hybrids}
S.~Huda, J.~Abawajy, M.~Alazab, M.~Abdollalihian, R.~Islam, J.~Yearwood,
  Hybrids of support vector machine wrapper and filter based framework for
  malware detection, Future Generation Computer Systems 55 (2016) 376--390.
\newblock \href {https://doi.org/https://doi.org/10.1016/j.future.2014.06.001}
  {\path{doi:https://doi.org/10.1016/j.future.2014.06.001}}.

\bibitem{5615097}
M.~Alazab, S.~Venkataraman, P.~Watters, Towards understanding malware behaviour
  by the extraction of api calls, in: 2010 Second Cybercrime and Trustworthy
  Computing Workshop, 2010, pp. 52--59.
\newblock \href {https://doi.org/10.1109/CTC.2010.8}
  {\path{doi:10.1109/CTC.2010.8}}.

\bibitem{Zelinka_Amer_2019}
I.~Zelinka, E.~Amer, An ensemble-based malware detection model using minimum
  feature set, MENDEL 25~(2) (2019) 1--10.
\newblock \href {https://doi.org/10.13164/mendel.2019.2.001}
  {\path{doi:10.13164/mendel.2019.2.001}}.

\bibitem{HAN2019208}
Maldae: Detecting and explaining malware based on correlation and fusion of
  static and dynamic characteristics, Computers \& Security 83 (2019) 208--233.
\newblock \href {https://doi.org/https://doi.org/10.1016/j.cose.2019.02.007}
  {\path{doi:https://doi.org/10.1016/j.cose.2019.02.007}}.

\bibitem{Obfuscat45:online}
Obfuscated files or information, technique t1027 - enterprise | mitre
  att\&ck®, \url{https://attack.mitre.org/techniques/T1027/}, (Accessed on
  07/08/2022).

\bibitem{udayakumar2017dynamic}
N.~Udayakumar, S.~Anandaselvi, T.~Subbulakshmi, Dynamic malware analysis using
  machine learning algorithm, in: 2017 International Conference on Intelligent
  Sustainable Systems (ICISS), IEEE, 2017, pp. 795--800.
\newblock \href {https://doi.org/https://doi.org/10.1109/ISS1.2017.8389286}
  {\path{doi:https://doi.org/10.1109/ISS1.2017.8389286}}.

\bibitem{Han2019a}
W.~Han, J.~Xue, Y.~Wang, Z.~Liu, Z.~Kong, {MalInsight: A systematic profiling
  based malware detection framework}, Journal of Network and Computer
  Applications 125 (2019) 236--250.
\newblock \href {https://doi.org/https://doi.org/10.1016/j.jnca.2018.10.022}
  {\path{doi:https://doi.org/10.1016/j.jnca.2018.10.022}}.

\bibitem{vemparala2019malware}
S.~Vemparala, F.~D. Troia, C.~A. Visaggio, T.~H. Austin, M.~Stamp, Malware
  detection using dynamic birthmarks (2019).
\newblock \href {http://arxiv.org/abs/1901.07312} {\path{arXiv:1901.07312}}.

\bibitem{8405026}
G.~Apruzzese, M.~Colajanni, L.~Ferretti, A.~Guido, M.~Marchetti, On the
  effectiveness of machine and deep learning for cyber security, in: 2018 10th
  International Conference on Cyber Conflict (CyCon), 2018, pp. 371--390.
\newblock \href {https://doi.org/10.23919/CYCON.2018.8405026}
  {\path{doi:10.23919/CYCON.2018.8405026}}.

\bibitem{Gibert2020}
D.~Gibert, C.~Mateu, J.~Planes, {The rise of machine learning for detection and
  classification of malware: Research developments, trends and challenges},
  Journal of Network and Computer Applications 153 (2020).
\newblock \href {https://doi.org/10.1016/j.jnca.2019.102526}
  {\path{doi:10.1016/j.jnca.2019.102526}}.

\bibitem{Bostami2020}
B.~Bostami, M.~Ahmed, Deep Learning Meets Malware Detection: An Investigation,
  2020, pp. 137--155.
\newblock \href {https://doi.org/10.1007/978-3-030-35642-2_7}
  {\path{doi:10.1007/978-3-030-35642-2_7}}.

\bibitem{li2022novel}
C.~Li, Q.~Lv, N.~Li, Y.~Wang, D.~Sun, Y.~Qiao, A novel deep framework for
  dynamic malware detection based on api sequence intrinsic features, Computers
  \& Security 116 (2022) 102686.

\bibitem{kim2018multimodal}
T.~Kim, B.~Kang, M.~Rho, S.~Sezer, E.~G. Im, A multimodal deep learning method
  for android malware detection using various features, IEEE Transactions on
  Information Forensics and Security 14~(3) (2018) 773--788.

\bibitem{catak2020deep}
F.~O. Catak, A.~F. Yaz{\i}, O.~Elezaj, J.~Ahmed, Deep learning based sequential
  model for malware analysis using windows exe api calls, PeerJ Computer
  Science 6 (2020) e285.
\newblock \href {https://doi.org/https://doi.org/10.7717/peerj-cs.285}
  {\path{doi:https://doi.org/10.7717/peerj-cs.285}}.

\bibitem{Tirumala2020}
S.~S. Tirumala, M.~R. Valluri, D.~Nanadigam, {Evaluation of Feature and
  Signature based Training Approaches for Malware Classification using
  Autoencoders}, in: 2020 International Conference on COMmunication Systems and
  NETworkS(COMSNETS), 2020, pp. 1--5.
\newblock \href
  {https://doi.org/https://doi.org/10.1109/COMSNETS48256.2020.9027373}
  {\path{doi:https://doi.org/10.1109/COMSNETS48256.2020.9027373}}.

\bibitem{8802378}
A.~Sharma, P.~Malacaria, M.~Khouzani, Malware detection using 1-dimensional
  convolutional neural networks, in: 2019 IEEE European Symposium on Security
  and Privacy Workshops (EuroS PW), 2019, pp. 247--256.

\bibitem{9204665}
R.~Feng, S.~Chen, X.~Xie, G.~Meng, S.-W. Lin, Y.~Liu, A performance-sensitive
  malware detection system using deep learning on mobile devices, IEEE
  Transactions on Information Forensics and Security 16 (2021) 1563--1578.
\newblock \href {https://doi.org/10.1109/TIFS.2020.3025436}
  {\path{doi:10.1109/TIFS.2020.3025436}}.

\bibitem{8556824}
K.~Alrawashdeh, C.~Purdy, Ransomware detection using limited precision deep
  learning structure in fpga, in: NAECON 2018 - IEEE National Aerospace and
  Electronics Conference, 2018, pp. 152--157.

\bibitem{Suaboot2020}
J.~Suaboot, Z.~Tari, A.~Mahmood, A.~Y. Zomaya, W.~Li, {Sub-curve HMM: A malware
  detection approach based on partial analysis of API call sequences},
  Computers {\&} Security 92 (2020) 1--15.
\newblock \href {https://doi.org/https://doi.org/10.1016/j.cose.2020.101773}
  {\path{doi:https://doi.org/10.1016/j.cose.2020.101773}}.

\bibitem{amer2022malware}
E.~Amer, A.~Samir, H.~Mostafa, A.~Mohamed, M.~Amin, Malware detection approach
  based on the swarm-based behavioural analysis over api calling sequence, in:
  2022 2nd International Mobile, Intelligent, and Ubiquitous Computing
  Conference (MIUCC), IEEE, 2022, pp. 27--32.

\bibitem{moraffah2020causal}
R.~Moraffah, M.~Karami, R.~Guo, A.~Raglin, H.~Liu, Causal interpretability for
  machine learning-problems, methods and evaluation, ACM SIGKDD Explorations
  Newsletter 22~(1) (2020) 18--33.
\newblock \href {https://doi.org/https://doi.org/10.1145/3400051.3400058}
  {\path{doi:https://doi.org/10.1145/3400051.3400058}}.

\bibitem{mehrabi2019survey}
N.~Mehrabi, F.~Morstatter, N.~Saxena, K.~Lerman, A.~Galstyan, A survey on bias
  and fairness in machine learning, arXiv preprint arXiv:1908.09635 (2019).

\bibitem{ribeiro2016should}
M.~T. Ribeiro, S.~Singh, C.~Guestrin, " why should i trust you?" explaining the
  predictions of any classifier, in: Proceedings of the 22nd ACM SIGKDD
  international conference on knowledge discovery and data mining, 2016, pp.
  1135--1144.

\bibitem{mimura2022applying}
M.~Mimura, R.~Ito, Applying nlp techniques to malware detection in a practical
  environment, International Journal of Information Security 21~(2) (2022)
  279--291.

\bibitem{Amer2020}
E.~Amer, I.~Zelinka, {A dynamic Windows malware detection and prediction method
  based on contextual understanding of API call sequence}, Computers {\&}
  Security 92 (2020).
\newblock \href {https://doi.org/https://doi.org/10.1016/j.cose.2020.101760}
  {\path{doi:https://doi.org/10.1016/j.cose.2020.101760}}.

\bibitem{Sihwail2019}
R.~Sihwail, K.~Omar, K.~A.~Z. Ariffin, S.~{Al Afghani}, {Malware detection
  approach based on artifacts in memory image and dynamic analysis}, Applied
  Sciences 9~(18) (2019).
\newblock \href {https://doi.org/https://doi.org/10.3390/app9183680}
  {\path{doi:https://doi.org/10.3390/app9183680}}.

\bibitem{Singh2020b}
J.~Singh, J.~Singh, {A survey on machine learning-based malware detection in
  executable files}, Journal of Systems Architecture (2020).
\newblock \href {https://doi.org/https://doi.org/10.1016/j.sysarc.2020.101861}
  {\path{doi:https://doi.org/10.1016/j.sysarc.2020.101861}}.

\bibitem{molnar2020interpretable}
C.~Molnar, Interpretable machine learning, Lulu. com, 2020.

\bibitem{karbab2019maldy}
E.~B. Karbab, M.~Debbabi, Maldy: Portable, data-driven malware detection using
  natural language processing and machine learning techniques on behavioral
  analysis reports, Digital Investigation 28 (2019) S77--S87.
\newblock \href {https://doi.org/https://doi.org/10.1016/j.diin.2019.01.017}
  {\path{doi:https://doi.org/10.1016/j.diin.2019.01.017}}.

\bibitem{Operatin76:online}
Operating system market share,
  \url{https://netmarketshare.com/operating-system-market-share.aspx?options},
  (Accessed on 11/04/2021).

\bibitem{Over100M74:online}
A.~Drapkin, Over 100 million pieces of malware were made for windows users in
  2021, \url{https://tech.co/news/windows-users-malware}, (Accessed on
  04/10/2022).

\bibitem{WindowsU4:online}
T.~Boris, Windows users beware: 95\% of ransomware attacks target microsoft’s
  os [google report] | tech times,
  \url{https://www.techtimes.com/articles/266728/20211015/windows-users-ransomware-attack-windows-ransomware-windows-microsoft-google-report.htm#:~:text=Windows%20users%20have%20been%20the,of%20Google%2C%20VirusTotal%2C%20reported.},
  (Accessed on 04/10/2022).

\bibitem{Quickint67:online}
M.~A. Stenne, Quick introduction to windows api,,
  \url{https://users.physics.ox.ac.uk/~Steane/cpp_help/winapi_intro.htm},
  (Accessed on 05/14/2021).

\bibitem{Silberschatz2018}
G.~B.~P. Silberschatz~Abraham, Gagne~Greg, {Operating System Concepts}, 10th
  Edition, Wiley, 2018.

\bibitem{Uppal2014}
D.~Uppal, R.~Sinha, V.~Mehra, V.~Jain, Exploring behavioral aspects of api
  calls for malware identification and categorization, in: 2014 International
  Conference on Computational Intelligence and Communication Networks, 2014,
  pp. 824--828.
\newblock \href {https://doi.org/10.1109/CICN.2014.176}
  {\path{doi:10.1109/CICN.2014.176}}.

\bibitem{Programm42:online}
Microsoft, Programming reference for the win32 api - win32 apps | microsoft
  docs, \url{https://docs.microsoft.com/en-us/windows/win32/api/}, (Accessed on
  06/01/2021).

\bibitem{Ammar206656}
B.~I. A.~B. Ammar Ahmed E.~Elhadi, Mohd Aizaini~Maarof, {Improving the
  Detection of Malware Behaviour Using Simplified Data Dependent API Call
  Graph}, {International Journal of Security and Its Applications} {7} (2013)
  29--42.

\bibitem{ki2015novel}
Y.~Ki, E.~Kim, H.~K. Kim, A novel approach to detect malware based on api call
  sequence analysis, International Journal of Distributed Sensor Networks
  11~(6) (2015).
\newblock \href {https://doi.org/https://doi.org/10.1155/2015/659101}
  {\path{doi:https://doi.org/10.1155/2015/659101}}.

\bibitem{gupta2016malware}
S.~Gupta, H.~Sharma, S.~Kaur, Malware characterization using windows api call
  sequences, in: International Conference on Security, Privacy, and Applied
  Cryptography Engineering, 2016, pp. 271--280.
\newblock \href {https://doi.org/https://doi.org/10.1007/978-3-319-49445-6_15}
  {\path{doi:https://doi.org/10.1007/978-3-319-49445-6_15}}.

\bibitem{zhao2019feature}
Y.~Zhao, B.~Bo, Y.~Feng, C.~Xu, B.~Yu, A feature extraction method of hybrid
  gram for malicious behavior based on machine learning, Security and
  Communication Networks 2019 (2019).
\newblock \href {https://doi.org/https://doi.org/10.1155/2019/2674684}
  {\path{doi:https://doi.org/10.1155/2019/2674684}}.

\bibitem{ALAEIYAN201976}
M.~Alaeiyan, S.~Parsa, M.~Conti,
  \href{https://www.sciencedirect.com/science/article/pii/S0140366418300410}{Analysis
  and classification of context-based malware behavior}, Computer
  Communications 136 (2019) 76--90.
\newblock \href {https://doi.org/https://doi.org/10.1016/j.comcom.2019.01.003}
  {\path{doi:https://doi.org/10.1016/j.comcom.2019.01.003}}.
\newline\urlprefix\url{https://www.sciencedirect.com/science/article/pii/S0140366418300410}

\bibitem{DING201873}
Y.~Ding, X.~Xia, S.~Chen, Y.~Li, A malware detection method based on family
  behavior graph, Computers \& Security 73 (2018) 73--86.
\newblock \href {https://doi.org/https://doi.org/10.1016/j.cose.2017.10.007}
  {\path{doi:https://doi.org/10.1016/j.cose.2017.10.007}}.

\bibitem{SHARMA202124}
N.~Sharma, R.~Sharma, N.~Jindal, Machine learning and deep learning
  applications-a vision, Global Transitions Proceedings 2~(1) (2021) 24--28,
  1st International Conference on Advances in Information, Computing and Trends
  in Data Engineering (AICDE - 2020).
\newblock \href {https://doi.org/https://doi.org/10.1016/j.gltp.2021.01.004}
  {\path{doi:https://doi.org/10.1016/j.gltp.2021.01.004}}.

\bibitem{YUAN2021102221}
S.~Yuan, X.~Wu, Deep learning for insider threat detection: Review, challenges
  and opportunities, Computers \& Security 104 (2021) 102221.
\newblock \href {https://doi.org/https://doi.org/10.1016/j.cose.2021.102221}
  {\path{doi:https://doi.org/10.1016/j.cose.2021.102221}}.

\bibitem{najafabadi2015deep}
M.~M. Najafabadi, F.~Villanustre, T.~M. Khoshgoftaar, N.~Seliya, R.~Wald,
  E.~Muharemagic, Deep learning applications and challenges in big data
  analytics, Journal of big data 2~(1) (2015) 1--21.
\newblock \href {https://doi.org/https://doi.org/10.1186/s40537-014-0007-7}
  {\path{doi:https://doi.org/10.1186/s40537-014-0007-7}}.

\bibitem{rafique2020malware}
M.~F. Rafique, M.~Ali, A.~S. Qureshi, A.~Khan, A.~M. Mirza, Malware
  classification using deep learning based feature extraction and wrapper based
  feature selection technique (2020).
\newblock \href {http://arxiv.org/abs/1910.10958} {\path{arXiv:1910.10958}}.

\bibitem{PINHERO2021102247}
A.~Pinhero, A.~{M L}, V.~P, C.~Visaggio, A.~N, A.~S, A.~S, Malware detection
  employed by visualization and deep neural network, Computers \& Security 105
  (2021) 102247.
\newblock \href {https://doi.org/https://doi.org/10.1016/j.cose.2021.102247}
  {\path{doi:https://doi.org/10.1016/j.cose.2021.102247}}.

\bibitem{hubel1968receptive}
D.~H. Hubel, T.~N. Wiesel, Receptive fields and functional architecture of
  monkey striate cortex, The Journal of physiology 195~(1) (1968) 215--243.

\bibitem{fukushima1979neural}
K.~Fukushima, Neural network model for a mechanism of pattern recognition
  unaffected by shift in position-neocognitron, IEICE Technical Report, A
  62~(10) (1979) 658--665.

\bibitem{khan2018guide}
S.~Khan, H.~Rahmani, S.~A.~A. Shah, M.~Bennamoun, A guide to convolutional
  neural networks for computer vision, Synthesis Lectures on Computer Vision
  8~(1) (2018) 1--207.

\bibitem{lundervold2019overview}
A.~S. Lundervold, A.~Lundervold, An overview of deep learning in medical
  imaging focusing on mri, Zeitschrift f{\"u}r Medizinische Physik 29~(2)
  (2019) 102--127.
\newblock \href {https://doi.org/https://doi.org/10.1016/j.zemedi.2018.11.002}
  {\path{doi:https://doi.org/10.1016/j.zemedi.2018.11.002}}.

\bibitem{o2015introduction}
K.~O'Shea, R.~Nash, An introduction to convolutional neural networks, arXiv
  preprint arXiv:1511.08458 (2015).

\bibitem{yamashita2018convolutional}
R.~Yamashita, M.~Nishio, R.~K.~G. Do, K.~Togashi, Convolutional neural
  networks: an overview and application in radiology, Insights into imaging
  9~(4) (2018) 611--629.

\bibitem{kiranyaz2015real}
S.~Kiranyaz, T.~Ince, M.~Gabbouj, Real-time patient-specific ecg classification
  by 1-d convolutional neural networks, IEEE Transactions on Biomedical
  Engineering 63~(3) (2015) 664--675.

\bibitem{kim2014convolutional}
Y.~Kim, Convolutional neural networks for sentence classification (2014).
\newblock \href {http://arxiv.org/abs/1408.5882} {\path{arXiv:1408.5882}}.

\bibitem{fesseha2021text}
A.~Fesseha, S.~Xiong, E.~D. Emiru, M.~Diallo, A.~Dahou, Text classification
  based on convolutional neural networks and word embedding for low-resource
  languages: Tigrinya, Information 12~(2) (2021) 52.

\bibitem{avci2018efficiency}
O.~Avci, O.~Abdeljaber, S.~Kiranyaz, B.~Boashash, H.~Sodano, D.~J. Inman,
  Efficiency validation of one dimensional convolutional neural networks for
  structural damage detection using a shm benchmark data, in: Proc. 25th Int.
  Conf. Sound Vib.(ICSV), 2018, pp. 4600--4607.

\bibitem{kiranyaz20211d}
S.~Kiranyaz, O.~Avci, O.~Abdeljaber, T.~Ince, M.~Gabbouj, D.~J. Inman, 1d
  convolutional neural networks and applications: A survey, Mechanical systems
  and signal processing 151 (2021) 107398.
\newblock \href {https://doi.org/https://doi.org/10.1016/j.ymssp.2020.107398}
  {\path{doi:https://doi.org/10.1016/j.ymssp.2020.107398}}.

\bibitem{WANG2016806}
P.~Wang, B.~Xu, J.~Xu, G.~Tian, C.-L. Liu, H.~Hao, Semantic expansion using
  word embedding clustering and convolutional neural network for improving
  short text classification, Neurocomputing 174 (2016) 806--814.
\newblock \href {https://doi.org/https://doi.org/10.1016/j.neucom.2015.09.096}
  {\path{doi:https://doi.org/10.1016/j.neucom.2015.09.096}}.

\bibitem{johnson2015effective}
R.~Johnson, T.~Zhang, Effective use of word order for text categorization with
  convolutional neural networks (2015).
\newblock \href {http://arxiv.org/abs/1412.1058} {\path{arXiv:1412.1058}}.

\bibitem{LIU202021}
F.~Liu, L.~Zheng, J.~Zheng, Hienn-dwe: A hierarchical neural network with
  dynamic word embeddings for document level sentiment classification,
  Neurocomputing 403 (2020) 21--32.
\newblock \href {https://doi.org/https://doi.org/10.1016/j.neucom.2020.04.084}
  {\path{doi:https://doi.org/10.1016/j.neucom.2020.04.084}}.

\bibitem{lynn2019deep}
H.~M. Lynn, S.~B. Pan, P.~Kim, A deep bidirectional gru network model for
  biometric electrocardiogram classification based on recurrent neural
  networks, IEEE Access 7 (2019) 145395--145405.

\bibitem{hochreiter1997long}
S.~Hochreiter, J.~Schmidhuber, Long short-term memory, Neural computation 9~(8)
  (1997) 1735--1780.

\bibitem{cho2014learning}
K.~Cho, B.~Van~Merri{\"e}nboer, C.~Gulcehre, D.~Bahdanau, F.~Bougares,
  H.~Schwenk, Y.~Bengio, Learning phrase representations using rnn
  encoder-decoder for statistical machine translation, arXiv preprint
  arXiv:1406.1078 (2014).

\bibitem{horn2012matrix}
R.~A. Horn, C.~R. Johnson, Matrix analysis, Cambridge university press, 2012.

\bibitem{chung2014empirical}
J.~Chung, C.~Gulcehre, K.~Cho, Y.~Bengio, Empirical evaluation of gated
  recurrent neural networks on sequence modeling, arXiv preprint
  arXiv:1412.3555 (2014).

\bibitem{vukotic2016step}
V.~Vukoti{\'c}, C.~Raymond, G.~Gravier, A step beyond local observations with a
  dialog aware bidirectional gru network for spoken language understanding, in:
  Interspeech, 2016.

\bibitem{huda2017fast}
S.~Huda, J.~Abawajy, M.~Abdollahian, R.~Islam, J.~Yearwood, A fast malware
  feature selection approach using a hybrid of multi-linear and stepwise binary
  logistic regression, Concurrency and Computation: Practice and Experience
  29~(23) (2017) e3912.
\newblock \href {https://doi.org/https://doi.org/10.1002/cpe.3912}
  {\path{doi:https://doi.org/10.1002/cpe.3912}}.

\bibitem{fan2016malicious}
Y.~Fan, Y.~Ye, L.~Chen, Malicious sequential pattern mining for automatic
  malware detection, Expert Systems with Applications 52 (2016) 16--25.
\newblock \href {https://doi.org/https://doi.org/10.1016/j.eswa.2016.01.002}
  {\path{doi:https://doi.org/10.1016/j.eswa.2016.01.002}}.

\bibitem{raff2017learning}
E.~Raff, J.~Sylvester, C.~Nicholas, Learning the pe header, malware detection
  with minimal domain knowledge, in: Proceedings of the 10th ACM Workshop on
  Artificial Intelligence and Security, 2017, pp. 121--132.

\bibitem{Gettings10:online}
Getting started, \url{https://keras.io/getting_started/}, (Accessed on
  07/24/2022).

\bibitem{d2022apicula}
M.~D’Onghia, M.~Salvadore, B.~M. Nespoli, M.~Carminati, M.~Polino, S.~Zanero,
  Ap{\'\i}cula: Static detection of api calls in generic streams of bytes,
  Computers \& Security (2022) 102775.

\bibitem{kundu2021empirical}
P.~P. Kundu, L.~Anatharaman, T.~Truong-Huu, An empirical evaluation of
  automated machine learning techniques for malware detection, in: Proceedings
  of the 2021 ACM Workshop on Security and Privacy Analytics, 2021, pp. 75--81.
\newblock \href {https://doi.org/https://doi.org/10.1145/3445970.3451155}
  {\path{doi:https://doi.org/10.1145/3445970.3451155}}.

\bibitem{kale2022malware}
A.~S. Kale, V.~Pandya, F.~Di~Troia, M.~Stamp, Malware classification with
  word2vec, hmm2vec, bert, and elmo, Journal of Computer Virology and Hacking
  Techniques (2022) 1--16.

\bibitem{yeboah2022nlp}
P.~N. Yeboah, H.~B. Baz~Musah, Nlp technique for malware detection using 1d cnn
  fusion model, Security and Communication Networks 2022 (2022).

\bibitem{safebreach2017}
A.~Klein, I.~Kotler, {The Adventures of AV and the Leaky Sandbox: A SafeBreach
  Labs Research}, Tech. rep. (2017).

\bibitem{mar4413009}
L.~Martignoni, M.~Christodorescu, S.~Jha, Omniunpack: Fast, generic, and safe
  unpacking of malware, in: Twenty-Third Annual Computer Security Applications
  Conference (ACSAC 2007), 2007, pp. 431--441.
\newblock \href {https://doi.org/10.1109/ACSAC.2007.15}
  {\path{doi:10.1109/ACSAC.2007.15}}.

\bibitem{Yucel2020}
{\c{C}}.~Y{\"{u}}cel, A.~Koltuksuz, {Imaging and evaluating the memory access
  for malware}, Forensic Science International: Digital Investigation 32
  (2020).
\newblock \href {https://doi.org/10.1016/j.fsidi.2019.200903}
  {\path{doi:10.1016/j.fsidi.2019.200903}}.

\bibitem{ye2017m}
Y.~Ye, T.~Li, D.~Adjeroh, S.~S. Iyengar, A survey on malware detection using
  data mining techniques, ACM Computing Surveys 50~(3) (2017) 1--40.
\newblock \href {https://doi.org/https://doi.org/10.1145/3073559}
  {\path{doi:https://doi.org/10.1145/3073559}}.

\bibitem{CountVe93:online}
S.~Saket, (9) count vectorizer vs tfidf vectorizer | natural language
  processing | linkedin,
  \url{https://www.linkedin.com/pulse/count-vectorizers-vs-tfidf-natural-language-processing-sheel-saket/},
  (Accessed on 10/03/2021).

\bibitem{mandelbaum2016word}
A.~Mandelbaum, A.~Shalev, Word embeddings and their use in sentence
  classification tasks (2016).
\newblock \href {http://arxiv.org/abs/1610.08229} {\path{arXiv:1610.08229}}.

\bibitem{Pirscoveanu2015}
R.~S. Pirscoveanu, S.~S. Hansen, T.~M.~T. Larsen, M.~Stevanovic, J.~M.
  Pedersen, A.~Czech, {Analysis of Malware Behavior: Type Classification using
  Machine Learning}, in: 2015 International Conference on Cyber Situational
  Awareness, Data Analytics and Assessment (CyberSA), 2015.
\newblock \href {https://doi.org/https://doi.org/10.1109/CyberSA.2015.7166115}
  {\path{doi:https://doi.org/10.1109/CyberSA.2015.7166115}}.

\bibitem{Pan2016}
Z.-P. Pan, C.~Feng, C.-J. Tang, {Malware Classification Based on the Behavior
  Analysis and Back Propagation Neural Network}, in: 3rd Annual International
  Conference on Information Technology and Applications (ITA 2016), Vol.~7,
  2016, pp. 1--5.
\newblock \href {https://doi.org/https://doi.org/10.1051/itmconf/20160702001}
  {\path{doi:https://doi.org/10.1051/itmconf/20160702001}}.

\bibitem{mira2016novel}
F.~Mira, A.~Brown, W.~Huang, Novel malware detection methods by using lcs and
  lcss, in: 2016 22nd International Conference on Automation and Computing
  (ICAC), 2016, pp. 554--559.
\newblock \href {https://doi.org/https://doi.org/10.1109/IConAC.2016.7604978}
  {\path{doi:https://doi.org/10.1109/IConAC.2016.7604978}}.

\bibitem{cho2016malware}
I.~K. Cho, T.~G. Kim, Y.~J. Shim, M.~Ryu, E.~G. Im, Malware analysis and
  classification using sequence alignments, Intelligent Automation \& Soft
  Computing 22~(3) (2016) 371--377.
\newblock \href {https://doi.org/https://doi.org/10.1080/10798587.2015.1118916}
  {\path{doi:https://doi.org/10.1080/10798587.2015.1118916}}.

\bibitem{stiborek2018multiple}
J.~Stiborek, T.~Pevny, M.~Rehak, Multiple instance learning for malware
  classification, Expert Systems with Applications 93 (2018) 346--357.
\newblock \href {https://doi.org/https://doi.org/10.1016/j.eswa.2017.10.036}
  {\path{doi:https://doi.org/10.1016/j.eswa.2017.10.036}}.

\bibitem{wucher7867799}
T.~Wüchner, A.~Cisłak, M.~Ochoa, A.~Pretschner, Leveraging compression-based
  graph mining for behavior-based malware detection, IEEE Transactions on
  Dependable and Secure Computing 16~(1) (2019) 99--112.
\newblock \href {https://doi.org/10.1109/TDSC.2017.2675881}
  {\path{doi:10.1109/TDSC.2017.2675881}}.

\bibitem{6703684}
S.~Jha, M.~Fredrikson, M.~Christodoresu, R.~Sailer, X.~Yan, Synthesizing
  near-optimal malware specifications from suspicious behaviors, in: 2013 8th
  International Conference on Malicious and Unwanted Software: "The Americas"
  (MALWARE), 2013, pp. 41--50.
\newblock \href {https://doi.org/10.1109/MALWARE.2013.6703684}
  {\path{doi:10.1109/MALWARE.2013.6703684}}.

\bibitem{6679854}
K.~Blokhin, J.~Saxe, D.~Mentis, Malware similarity identification using call
  graph based system call subsequence features, in: 2013 IEEE 33rd
  International Conference on Distributed Computing Systems Workshops, 2013,
  pp. 6--10.
\newblock \href {https://doi.org/10.1109/ICDCSW.2013.55}
  {\path{doi:10.1109/ICDCSW.2013.55}}.

\bibitem{9318301}
A.~Hellal, F.~Mallouli, A.~Hidri, R.~K. Aljamaeen, A survey on graph-based
  methods for malware detection, in: 2020 4th International Conference on
  Advanced Systems and Emergent Technologies, 2020, pp. 130--134.
\newblock \href {https://doi.org/10.1109/IC_ASET49463.2020.9318301}
  {\path{doi:10.1109/IC_ASET49463.2020.9318301}}.

\bibitem{ding2018malware}
Y.~Ding, X.~Xia, S.~Chen, Y.~Li, A malware detection method based on family
  behavior graph, Computers \& Security 73 (2018) 73--86.
\newblock \href {https://doi.org/https://doi.org/10.1016/j.cose.2017.10.007}
  {\path{doi:https://doi.org/10.1016/j.cose.2017.10.007}}.

\bibitem{pei2020amalnet}
X.~Pei, L.~Yu, S.~Tian, Amalnet: A deep learning framework based on graph
  convolutional networks for malware detection, Computers \& Security 93 (2020)
  101792.
\newblock \href {https://doi.org/https://doi.org/10.1016/j.cose.2020.101792}
  {\path{doi:https://doi.org/10.1016/j.cose.2020.101792}}.

\bibitem{8233569}
T.~K. Tran, H.~Sato, Nlp-based approaches for malware classification from api
  sequences, in: 2017 21st Asia Pacific Symposium on Intelligent and
  Evolutionary Systems (IES), 2017, pp. 101--105.
\newblock \href {https://doi.org/10.1109/IESYS.2017.8233569}
  {\path{doi:10.1109/IESYS.2017.8233569}}.

\bibitem{liu2019robust}
Y.~Liu, Y.~Wang, A robust malware detection system using deep learning on api
  calls, in: 2019 IEEE 3rd Information Technology, Networking, Electronic and
  Automation Control Conference (ITNEC), IEEE, 2019, pp. 1456--1460.

\bibitem{li2022intelligent}
S.~Li, Q.~Zhou, R.~Zhou, Q.~Lv, Intelligent malware detection based on graph
  convolutional network, The Journal of Supercomputing 78~(3) (2022)
  4182--4198.

\bibitem{lajevardi2022markhor}
A.~M. Lajevardi, S.~Parsa, M.~J. Amiri, Markhor: malware detection using fuzzy
  similarity of system call dependency sequences, Journal of Computer Virology
  and Hacking Techniques 18~(2) (2022) 81--90.

\bibitem{8358312}
S.~Maniath, A.~Ashok, P.~Poornachandran, V.~Sujadevi, P.~Sankar~A.U., S.~Jan,
  Deep learning lstm based ransomware detection, in: 2017 Recent Developments
  in Control, Automation \& Power Engineering (RDCAPE), 2017, pp. 442--446.

\bibitem{chen2022cruparamer}
X.~Chen, Z.~Hao, L.~Li, L.~Cui, Y.~Zhu, Z.~Ding, Y.~Liu, Cruparamer: Learning
  on parameter-augmented api sequences for malware detection, IEEE Transactions
  on Information Forensics and Security 17 (2022) 788--803.

\bibitem{9777099}
M.~Sukul, S.~A. Lakshmanan, R.~Gowtham, Automated dynamic detection of
  ransomware using augmented bootstrapping, in: 2022 6th International
  Conference on Trends in Electronics and Informatics (ICOEI), 2022, pp.
  787--794.

\bibitem{dhanya2022performance}
L.~Dhanya, R.~Chitra, A.~A. Bamini, Performance evaluation of various ensemble
  classifiers for malware detection, Materials Today: Proceedings (2022).

\bibitem{abbasi2022behavior}
M.~S. Abbasi, H.~Al-Sahaf, M.~Mansoori, I.~Welch, Behavior-based ransomware
  classification: A particle swarm optimization wrapper-based approach for
  feature selection, Applied Soft Computing 121 (2022) 108744.

\bibitem{jing2022ensemble}
C.~Jing, Y.~Wu, C.~Cui, Ensemble dynamic behavior detection method for
  adversarial malware, Future Generation Computer Systems 130 (2022) 193--206.

\bibitem{PDFWindo91:online}
N.~Allan, J.~Ngubiri, Windows pe api calls for malicious and benigin programs,
  \url{https://www.researchgate.net/publication/336024802_Windows_PE_API_calls_for_malicious_and_benigin_programs?channel=doi\&linkId=5d8b6a4e92851c33e9395c89\&showFulltext=true},
  (Accessed on 12/16/2021).

\bibitem{ceschin2018need}
F.~Ceschin, F.~Pinage, M.~Castilho, D.~Menotti, L.~S. Oliveira, A.~Gregio, The
  need for speed: An analysis of brazilian malware classifiers, IEEE Security
  \& Privacy 16~(6) (2018) 31--41.

\bibitem{GitHuble77:online}
Github - leocsato/detector\_mw: Optimizer for malware detection. api calls
  sequence of benign files are provided.,
  \url{https://github.com/leocsato/detector_mw}, (Accessed on 12/15/2021).

\bibitem{mpascoMa95:online}
P.~Maniriho, Malbehavd-v1: A new dataset of api calls extracted from windows pe
  files of benign and malware, \url{https://github.com/mpasco/MalbehavD-V1},
  (Accessed on 04/07/2022).

\bibitem{rieck2011automatic}
K.~Rieck, P.~Trinius, C.~Willems, T.~Holz, Automatic analysis of malware
  behavior using machine learning, Journal of Computer Security 19~(4) (2011)
  639--668.
\newblock \href {https://doi.org/DOI: 10.3233/JCS-2010-0410} {\path{doi:DOI:
  10.3233/JCS-2010-0410}}.

\bibitem{Sethi2017}
K.~Sethi, B.~K. Tripathy, S.~K. Chaudhary, P.~Bera, {A Novel Malware Analysis
  for Malware Detection and Classification using Machine Learning Algorithms},
  in: SIN '17: Proceedings of the 10th International Conference on Security of
  Information and Networks, 2017, pp. 107--116.
\newblock \href {https://doi.org/https://doi.org/10.1145/3136825.3136883}
  {\path{doi:https://doi.org/10.1145/3136825.3136883}}.

\bibitem{nair2010medusa}
V.~P. Nair, H.~Jain, Y.~K. Golecha, M.~S. Gaur, V.~Laxmi, Medusa: Metamorphic
  malware dynamic analysis usingsignature from api, in: Proceedings of the 3rd
  International Conference on Security of Information and Networks, 2010, pp.
  263--269.
\newblock \href {https://doi.org/https://doi.org/10.1145/1854099.1854152}
  {\path{doi:https://doi.org/10.1145/1854099.1854152}}.

\bibitem{nappa2015malicia}
A.~Nappa, M.~Z. Rafique, J.~Caballero, The malicia dataset: identification and
  analysis of drive-by download operations, International Journal of
  Information Security 14~(1) (2015) 15--33.
\newblock \href {https://doi.org/https://doi.org/10.1007/s10207-014-0248-7}
  {\path{doi:https://doi.org/10.1007/s10207-014-0248-7}}.

\bibitem{VirusTot34:online}
Virustotal - home, \url{https://www.virustotal.com/gui/home/upload}, (Accessed
  on 09/29/2021).

\bibitem{FreeSoft10:online}
Free software downloads and reviews for windows, android, mac, and ios – cnet
  download, \url{https://download.cnet.com/}, (Accessed on 12/15/2021).

\bibitem{VirusTot60:online}
Virustotal - home, \url{https://www.virustotal.com/gui/home/upload}, (Accessed
  on 08/04/2022).

\bibitem{CuckooSa69:online}
Cuckoo sandbox - automated malware analysis, \url{https://cuckoosandbox.org/},
  (Accessed on 05/14/2021).

\bibitem{birunda2021review}
S.~S. Birunda, R.~K. Devi, A review on word embedding techniques for text
  classification, in: Innovative Data Communication Technologies and
  Application, Springer, 2021, pp. 267--281.
\newblock \href {https://doi.org/https://doi.org/10.1007/978-981-15-9651-3_23}
  {\path{doi:https://doi.org/10.1007/978-981-15-9651-3_23}}.

\bibitem{rong2016word2vec}
X.~Rong, word2vec parameter learning explained (2016).
\newblock \href {http://arxiv.org/abs/1411.2738} {\path{arXiv:1411.2738}}.

\bibitem{mikolov2013efficient}
T.~Mikolov, K.~Chen, G.~Corrado, J.~Dean, Efficient estimation of word
  representations in vector space (2013).
\newblock \href {http://arxiv.org/abs/1301.3781} {\path{arXiv:1301.3781}}.

\bibitem{chandak2021}
A.~Chandak, W.~Lee, M.~Stamp, A comparison of word2vec, hmm2vec, and pca2vec
  for malware classification (2021).
\newblock \href {http://arxiv.org/abs/2103.05763} {\path{arXiv:2103.05763}}.

\bibitem{RecentTr81:online}
K.~Sarkar, Recent trends in natural language processing using deep learning |
  medium,
  \url{https://medium.com/@kanchansarkar/recent-trends-in-natural-language-processing-using-deep
  -learning-a1469fbd2ef}, (Accessed on 05/20/2021) (2017).

\bibitem{pittaras_2020}
N.~Pittaras, G.~Giannakopoulos, G.~Papadakis, V.~Karkaletsis, Text
  classification with semantically enriched word embeddings, Natural Language
  Engineering (2020) 1–35\href {https://doi.org/10.1017/S1351324920000170}
  {\path{doi:10.1017/S1351324920000170}}.

\bibitem{MUHAMMAD2021728}
P.~F. Muhammad, R.~Kusumaningrum, A.~Wibowo, Sentiment analysis using word2vec
  and long short-term memory (lstm) for indonesian hotel reviews, Procedia
  Computer Science 179 (2021) 728--735, 5th International Conference on
  Computer Science and Computational Intelligence 2020.
\newblock \href {https://doi.org/https://doi.org/10.1016/j.procs.2021.01.061}
  {\path{doi:https://doi.org/10.1016/j.procs.2021.01.061}}.

\bibitem{NAWANGSARI2019360}
R.~P. Nawangsari, R.~Kusumaningrum, A.~Wibowo, Word2vec for indonesian
  sentiment analysis towards hotel reviews: An evaluation study, Procedia
  Computer Science 157 (2019) 360--366, the 4th International Conference on
  Computer Science and Computational Intelligence (ICCSCI 2019) : Enabling
  Collaboration to Escalate Impact of Research Results for Society.
\newblock \href {https://doi.org/https://doi.org/10.1016/j.procs.2019.08.178}
  {\path{doi:https://doi.org/10.1016/j.procs.2019.08.178}}.

\bibitem{STEIN2019216}
R.~A. Stein, P.~A. Jaques, J.~F. Valiati, An analysis of hierarchical text
  classification using word embeddings, Information Sciences 471 (2019)
  216--232.
\newblock \href {https://doi.org/https://doi.org/10.1016/j.ins.2018.09.001}
  {\path{doi:https://doi.org/10.1016/j.ins.2018.09.001}}.

\bibitem{LOPEZ2020103823}
W.~López, J.~Merlino, P.~Rodríguez-Bocca, Learning semantic information from
  internet domain names using word embeddings, Engineering Applications of
  Artificial Intelligence 94 (2020) 103823.
\newblock \href
  {https://doi.org/https://doi.org/10.1016/j.engappai.2020.103823}
  {\path{doi:https://doi.org/10.1016/j.engappai.2020.103823}}.

\bibitem{ENRIQUEZ20161}
F.~Enríquez, J.~A. Troyano, T.~López-Solaz, An approach to the use of word
  embeddings in an opinion classification task, Expert Systems with
  Applications 66 (2016) 1--6.
\newblock \href {https://doi.org/https://doi.org/10.1016/j.eswa.2016.09.005}
  {\path{doi:https://doi.org/10.1016/j.eswa.2016.09.005}}.

\bibitem{pennington2014glove}
J.~Pennington, R.~Socher, C.~D. Manning, Glove: Global vectors for word
  representation, in: Proceedings of the 2014 conference on empirical methods
  in natural language processing (EMNLP), 2014, pp. 1532--1543.

\bibitem{brownlee2017deep}
J.~Brownlee, Deep learning for natural language processing: develop deep
  learning models for your natural language problems, 2017.

\bibitem{Embeddin90:online}
Embedding layer,
  \url{https://keras.io/api/layers/core_layers/embedding/#embedding-layer},
  (Accessed on 12/14/2021).

\bibitem{collobert2011natural}
R.~Collobert, J.~Weston, L.~Bottou, M.~Karlen, K.~Kavukcuoglu, P.~Kuksa,
  Natural language processing (almost) from scratch, Journal of machine
  learning research 12~(ARTICLE) (2011) 2493--2537.

\bibitem{srivastava2014dropout}
N.~Srivastava, G.~Hinton, A.~Krizhevsky, I.~Sutskever, R.~Salakhutdinov,
  Dropout: a simple way to prevent neural networks from overfitting, The
  journal of machine learning research 15~(1) (2014) 1929--1958.

\bibitem{jadon2020survey}
S.~Jadon, A survey of loss functions for semantic segmentation, in: 2020 IEEE
  Conference on Computational Intelligence in Bioinformatics and Computational
  Biology (CIBCB), IEEE, 2020, pp. 1--7.

\bibitem{kingma2014adam}
D.~P. Kingma, J.~Ba, Adam: A method for stochastic optimization, arXiv preprint
  arXiv:1412.6980 (2014).

\bibitem{yaqub2020state}
M.~Yaqub, J.~Feng, M.~S. Zia, K.~Arshid, K.~Jia, Z.~U. Rehman, A.~Mehmood,
  State-of-the-art cnn optimizer for brain tumor segmentation in magnetic
  resonance images, Brain Sciences 10~(7) (2020) 427.

\bibitem{GitHubfa61:online}
Github - fabriciojoc/brazilian-malware-dataset: Dataset containing thousands of
  malware and goodware collected in the brazilian cyberspace over years.,
  \url{https://github.com/fabriciojoc/brazilian-malware-dataset}, (Accessed on
  12/16/2021).

\bibitem{The95:online}
Pypi, Pypi · the python package index, \url{https://pypi.org/}, (Accessed on
  08/22/2021).

\bibitem{8636415}
F.~Ceschin, F.~Pinage, M.~Castilho, D.~Menotti, L.~S. Oliveira, A.~Gregio, The
  need for speed: An analysis of brazilian malware classifiers, IEEE Security
  Privacy 16~(6) (2018) 31--41.
\newblock \href {https://doi.org/10.1109/MSEC.2018.2875369}
  {\path{doi:10.1109/MSEC.2018.2875369}}.

\bibitem{Understa80:online}
L.~Hulstaert, Understanding model predictions with lime | by lars hulstaert |
  towards data science,
  \url{https://towardsdatascience.com/understanding-model-predictions-with-lime-a582fdff3a3b},
  (Accessed on 07/03/2022).

\bibitem{anchors:aaai18}
M.~T. Ribeiro, S.~Singh, C.~Guestrin, Anchors: High-precision model-agnostic
  explanations, in: AAAI Conference on Artificial Intelligence (AAAI), 2018.

\bibitem{GitHubTe20:online}
MIT, Github - teamhg-memex/eli5: A library for debugging/inspecting machine
  learning classifiers and explaining their predictions,
  \url{https://github.com/TeamHG-Memex/eli5}, (Accessed on 07/16/2022).

\end{thebibliography}

\end{document}